\documentclass[onecolumn,useAMS,usenatbib]{mn2e}
\usepackage{amsfonts}
\usepackage{graphicx}
\usepackage{bm}
\usepackage{psfig}

\newcommand{\Tsapprox}{|_{\scriptscriptstyle T_{s}}}
\newcommand{\Tseff}{T_{s}^{\scriptscriptstyle \rm eff}}

\newcommand{\rmd}{{\rm d}}
\newcommand{\rme}{{\rm e}}
\newcommand{\rmi}{{\rm i}}

\newcommand{\rmH}{{\rm H}}
\newcommand{\rmHe}{{\rm He}}

\newcommand{\HI}{$\rmH\,${\sc i}}

\newcommand{\vpec}{v^{({\rm pec})}}

\newcommand{\nHI}{n_{\rm HI}}
\newcommand{\nHeI}{n_{\rm HeI}}

\newcommand{\mH}{m_{\rm H}}
\newcommand{\mHe}{m_{\rm He}}
\newcommand{\lya}{{\ensuremath{{\rm Ly}\alpha}}}
\newcommand{\rad}{{\ensuremath{\rm rad}}}
\newcommand{\col}{{\ensuremath{\rm col}}}
\newcommand{\rth}{{\ensuremath{\rm th}}}

\newcommand{\bM}{{\bmath M}}
\newcommand{\be}{{\bmath e}}
\newcommand{\br}{{\bmath r}}
\newcommand{\bu}{{\bmath u}}
\newcommand{\bv}{{\bmath v}}
\newcommand{\bw}{{\bmath w}}
\newcommand{\bk}{{\bmath k}}
\newcommand{\bR}{{\bmath R}}
\newcommand{\bOmega}{{\bmath\Omega}}

\newcommand{\mtt}[9]{\left(\begin{array}{ccc} #1 & #2 & #3 \\ #4 & #5 & #6 
\\ #7 & #8 & #9 \end{array}\right)}

\newcommand{\fzz}{f_{00}(\hat\bOmega)}
\newcommand{\fzo}{f_{01}(\hat\bOmega)}
\newcommand{\foz}{f_{10}(\hat\bOmega)}
\newcommand{\foo}{f_{11}(\hat\bOmega)}

\newcommand{\beq}{\begin{equation}}
\newcommand{\eeq}{\end{equation}}
\newcommand{\beqa}{\begin{eqnarray}}
\newcommand{\eeqa}{\end{eqnarray}}

\title[Spin-resolved atomic velocity distribution]
{The spin-resolved atomic velocity distribution and 21-cm line profile of 
  dark-age gas}

\author[Hirata \& Sigurdson]
{Christopher M. Hirata\thanks{E-mail: {\tt chirata@sns.ias.edu} (CMH); 
  {\tt krs@sns.ias.edu} (KS)} and 
  Kris Sigurdson\footnotemark[1]\thanks{Hubble Fellow}
\\Institute for Advanced Study, Einstein Drive,
      Princeton, NJ 08540, USA
}

\date{\today}

\begin{document}
\maketitle

\begin{abstract}
The 21-cm hyperfine line of atomic hydrogen (\HI) is a promising probe of 
the cosmic dark ages.  In past treatments of 21-cm radiation it was 
assumed the hyperfine level populations of \HI\ could be characterized by 
a velocity-independent ``spin temperature'' $T_s$ determined by a 
competition between 21-cm radiative transitions, spin-changing collisions, 
and (at lower redshifts) \lya\ scattering.  However we show here that, if 
the collisional time is comparable to the radiative time, the spin 
temperature will depend on atomic velocity, $T_s=T_s(v)$, and one must 
replace the usual hyperfine level rate equations with a Boltzmann equation 
describing the spin and velocity dependence of the \HI\ distribution 
function.  We construct here the Boltzmann equation relevant to the cosmic 
dark ages and solve it using a basis-function method.  Accounting for the 
actual spin-resolved atomic velocity distribution results in up to a 
$\sim\!\!2$ per cent suppression of the 21-cm emissivity, and a redshift 
and angular-projection dependent suppression or enhancement of the linear 
power spectrum of 21-cm fluctuations of up to $\sim\!5$ per cent.  The 
effect on the 21-cm line profile is more dramatic --- its full-width at 
half maximum (FWHM) can be enhanced by up to $ \sim\!60$ per cent relative 
to the velocity-independent calculation.  We discuss the implications for 
21-cm tomography of the dark ages.
\end{abstract}

\begin{keywords}
cosmology: theory -- intergalactic medium -- atomic processes --
line: profiles.
\end{keywords}

\section{Introduction}

While both earlier and later epochs of the Universe have now been observed 
by astronomers, the cosmic dark ages (the time between the formation of 
the first atoms at $z\sim 1000$ and the subsequent formation of the first 
luminous objects) remain an observational enigma.  The difficulty of 
probing this epoch stems from the relatively inert physical properties 
expected of the dark-age gas. It is diffuse, without the luminous galaxies 
or quasars that have proven so useful for lower-redshift astrophysics. It 
is mostly neutral, hence signatures such as recombination lines, free-free 
radiation, and Compton scattering of the cosmic microwave background (CMB) 
are essentially absent.  Finally, its composition consists mainly of cold 
H and He atoms, which have very few lines that can be excited at the 
relevant temperatures.

A notable exception to this last property is the hyperfine 21-cm line of 
neutral hydrogen (\HI).  The \HI\ 21-cm line has attracted considerable 
attention as a possible probe of cosmic evolution during the dark ages, 
despite substantial observational challenges.  In particular by using 
frequency as well as angular information, it should be possible to map out 
the 21-cm signal in three dimensions, thereby greatly increasing the 
number of modes that can be observed \citep{2004PhRvL..92u1301L, 
2005astro.ph.10027M, 2005astro.ph.12262B}.  
Furthermore, the foregrounds to this signal are expected to have a 
distinctly different character in both angular and especially frequency 
space (they are expected to be relatively smooth in frequency space) which 
should allow for effective foreground removal \citep{2004ApJ...608..622Z, 
2005ApJ...625..575S, 2005astro.ph..1081W}.

The nature of the 21-cm signal depends on both the properties of the 
primordial gas and the mechanisms that excite or de-excite the hydrogen 
atoms.  The major mechanisms operating during the dark ages are the 
radiative transitions at $\lambda_{21}=21.1\,$cm and atom-atom collisions; 
the former tend to cause the hyperfine level populations to relax to the 
CMB temperature $T_\gamma$, whereas the latter tend to drive the level 
populations towards the gas kinetic temperature $T_k$.  The atomic gas is 
expected to be colder than the CMB during the dark ages since it cools 
adiabatically as $T\propto a^{-2}$ after residual Compton heating becomes 
ineffective.  During this period we thus expect to see the 21-cm signal in 
absorption when the collisional coupling is strong and 
\citet{2004PhRvL..92u1301L} have shown that at high redshifts $z\ge 50$ 
the entire Universe should be collisionally coupled and an absorption 
signal should be present.  At lower redshifts collisions become less 
efficient due to the low density and the level populations approach 
$T_\gamma$; in this case the absorption signal fades because the \HI\ 
spins are nearly in thermal equilibrium with the CMB.  Only the 
highest-density regions remain collisionally coupled, and these may even 
appear in emission if shocks or adiabatic compression heat them to 
$T_k>T_\gamma$ \citep{2002ApJ...572L.123I,2003MNRAS.341...81I}.  At later 
times when sources 
of ultraviolet radiation have turned on, scattering of photons in 
the \HI\ \lya\ resonance can also become significant, and can once again 
drive the level populations towards $T_k$ even in mean-density and 
underdense regions.  These early sources of radiation may also have 
emitted X-rays, which could heat the intergalactic medium (IGM) to well 
above the CMB temperature \citep{1997ApJ...475..429M, 2004ApJ...602....1C, 
2005MNRAS.363.1069K}, and perhaps produce sufficient ionization that 
atom-electron collisions are important in determining the spin temperature 
in some regions \citep{2006ApJ...637L...1K}.

A detailed understanding of all these excitation and de-excitation 
mechanisms and their effect on the 21-cm emissivity and line profile will 
be necessary in order to interpret future 21-cm observations.  In the 
1950s and 1960s, analysis of 21-cm radiation from the interstellar medium 
motivated pioneering studies of the \lya\ scattering 
\citep{1952AJ.....57R..31W,field1958,1959ApJ...129..536F} and collisional 
\citep{1961PRSLA.262..132D,1969ApJ...158..423A} mechanisms.  The results 
of these analyses have largely been carried over into the theory of 21-cm 
radiation from the high-redshift IGM and dark ages, with some improvements 
and adaptations to accommodate different physical situations.  For 
example, several authors have computed the efficiency of the \lya\ 
scattering mechanism with detailed consideration of cosmological radiative 
transfer effects, interaction with Ly$\beta$ and higher-order \HI\ Lyman 
lines, and the fine and hyperfine structure of the \lya\ resonance 
\citep{2004ApJ...602....1C, 2005astro.ph..7102H, 2005astro.ph..8381P, 
2005astro.ph.12206C}.  \citet{2005ApJ...622.1356Z} has also re-evaluated 
the collisional hyperfine excitation rate using modern atomic physics 
techniques.

In this paper we note that heretofore all studies of high-redshift 21-cm 
radiation have assumed that the hyperfine level population ratios of the 
hydrogen atoms in the primordial gas are independent of velocity.  If the 
level populations are also isotropic, they dependent only on the total 
spin $F$ and not on the magnetic quantum number $M_F$, so they can 
be characterized by a single spin temperature $T_s$.\footnote{Several 
groups have discussed the possibility of $M_F$-dependent populations due 
to magnetic fields \citep{2005MNRAS.359L..47C}, CMB anisotropy 
\citep{2005ApJ...622.1356Z}, and anisotropies in the \lya\ radiation 
\citep{2005ApJ...635....1B}.  In all cases the effect was argued to be 
very small, and indeed \citet{2005ApJ...622.1356Z} showed that if any net 
spin polarization of the \HI\ atoms was initially present, it would be 
destroyed by radiative transitions much faster than the Hubble time.} This 
assumption implies that the ratio of hyperfine level populations is 
$\nHI(F=1)/\nHI(F=0)\equiv 3\rme^{-T_\star/T_s}$ independent of the 
velocity of the hydrogen atoms.\footnote{Here $T_\star \equiv h 
c/\lambda_{21}k_B = 68.2$~mK is the energy of the 21-cm transition in 
temperature units.} This should be the case if the level populations are 
determined solely by radiative transitions with the CMB in the 21-cm line 
or solely by atomic collisions as, in these cases, the hyperfine level 
populations will thermalize at a temperature $T_\gamma$ or $T_k$ 
respectively. However we demonstrate here that, because the level 
populations are determined by both radiative and collisional transitions 
and the collisional spin-change cross section for H-H collisions is 
velocity dependent while the radiative cross section is velocity 
independent, \emph{the hyperfine level-resolved distribution function of 
\HI\ is not thermal}.  This is our main result and we show below that this 
has a significant (order unity) effect on the 21-cm line profile of 
collisionally coupled gas at high redshift.

The paper is organized as follows: We outline the basic kinetic theory and 
derive the collisional piece of the linear Boltzmann equation in 
Section~\ref{sec:kinetic} and define the basis-function method we use to 
describe deviations from a thermal distribution function in 
Section~\ref{sec:basis}.  In Section~\ref{sec:hhcoll} we evaluate the H-H 
collision term and in Section~\ref{sec:he} we evaluate the He-H collision 
term to account for the effects of helium.  We solve the Boltzmann 
equation in Section~\ref{sec:solveboltz} and present our results for the 
21-cm line profile in Section~\ref{sec:results}. A short discussion and 
summary of our main results follows in Section~\ref{sec:dis}.  We have 
included in Appendix~\ref{app:basis} a summary of some useful properties 
of the Gauss-Hermite basis functions, and in Appendix~\ref{app:cross} the 
details of the quantum mechanical calculation of the relevant atomic cross 
sections.  When needed for our numerical results we have assumed a flat 
cosmology with matter density $\Omega_m=0.30$, baryon density 
$\Omega_b=0.042$, Hubble constant $H_0=72\,$km$\,$s$^{-1}\,$Mpc$^{-1}$, 
primordial helium fraction $Y_{\rm He}=0.24$, and CMB temperature 
$T_0=2.728\,$K.

\section{Kinetic theory}
\label{sec:kinetic}

In the standard calculation of the 21-cm signal one uses the populations 
$\nHI(F=0)$ and $\nHI(F=1)$ of the ground and hyperfine-excited levels as 
the basic quantities which describe the state of the hydrogen atoms.  In 
this paper we aim to go beyond this calculation by solving for 
the joint distribution of spins and velocities.  This will be represented 
by the distribution function, $f_F(\bv)$, which describes the number of 
atoms in hyperfine level $F$ per unit volume in position space, 
per unit volume in velocity space.  For comparison, the standard 
calculation assumes that the distribution function can be 
separated into its kinetic degrees of freedom with temperature $T_k$ and 
its spin degrees of freedom with temperature $T_s$.  In this case one 
can write
\beq
f_F(\bv)\Tsapprox = \nHI y_F \Phi_{T_k}(\bv),
\label{eq:maxwellian}
\eeq
where $\Phi_{T_k}(\bv)$ represents the Maxwellian distribution,
\beq
\Phi_{T_k}(\bv) = \frac{1}{(2\pi\sigma^2)^{3/2}}
  \rme^{-v^2/2\sigma^2},
\eeq
$\sigma=\sqrt{k_BT_k/\mH}$ is the thermal velocity dispersion of the 
atoms, and $\bv$ is the velocity of the atom in the reference frame of the baryonic 
fluid.  Note that here and elsewhere in the paper the notation ``$|_{T_{s}}$'' is shorthand for ``evaluated in the spin-temperature approximation.'' The factor $y_F$ in Eq.~(\ref{eq:maxwellian})
is the fraction of the atoms in the $F$ hyperfine level, and an elementary calculation shows that this quantity is related to the spin 
temperature by
\beq
y_0 = \frac{1}{3\rme^{-T_\star/T_s}+1} \approx 
\frac14+\frac{3T_\star}{16T_s}
 \quad{\rm and}\quad
y_1 = \frac{3\rme^{-T_\star/T_s}}{3\rme^{-T_\star/T_s}+1} \approx
\frac34-\frac{3T_\star}{16T_s},
\label{eq:34}
\eeq
where the approximation $T_\star\ll T_s$ is valid in all cases of 
astrophysical interest.
The standard analysis solves for $T_s$ using the continuity equation for 
$F=0$ and $F=1$ \HI\ atoms.  By determining all rates (collisional, 
radiative, and \lya) of production or destruction of  \HI\ atoms in a given level one obtains the abundances $n(F)$ and hence the spin temperature 
$\rme^{-T_\star/T_s} = n(1)/[3n(0)]$.  In the standard case, the 
brightness temperature due to the 21-cm line is
\beq
T_b\Tsapprox = \left[
\frac{H(z)}{1+z} + \frac{\rmd \vpec_\parallel}{\rmd r_\parallel}
 \right]^{-1}
\frac{3c^3AT_\star}{32\pi\nu_{10}^3(1+z)^2}\nHI
\left(1-\frac{T_\gamma}{T_s}\right),
\label{eq:dtb-old}
\eeq
where $H(z)$ is the Hubble rate, $\vpec_\parallel$ is the radial component 
of the peculiar velocity, $c$ is the speed of light, $A$ is the 21-cm 
Einstein coefficient, $\nu_{10}$ is the frequency of the 21-cm line, and 
$T_\gamma$ is the CMB temperature \citep{2004MNRAS.352..142B, 
2005ApJ...624L..65B}.  One of the major goals of this paper is to 
update Eq.~(\ref{eq:dtb-old}) to include the full spin-velocity distribution.

\subsection{Simplifying approximations; distribution function}

The distribution function $f_F(\bv)$ is a much more general description of 
the state of the gas than the three numbers $\{\nHI,T_k,T_s\}$ in the 
standard treatment.  Nevertheless, the most general description possible 
would also account for dependence on $M_F$ and quantum correlations among 
the various states of \HI.  In this section we describe the key 
simplifying approximations that allow us to fully characterize the \HI\ 
atoms by a distribution function.

Since the spin state of a hydrogen atom is described by its two quantum 
numbers $|FM_F\rangle$, the most general characterization would be to 
write a quantum-mechanical density matrix $\rho_{FM_F,F'M'_F}(\br, \bv)$, 
so that e.g. $\rho_{00,00}(\br,\bv)$ would describe the number of H atoms 
in the ground state per unit volume, per unit volume in velocity space.  
The use of the density matrix allows for correlations between states of 
different $F$ or $M_F$, just as the $U$ and $V$ Stokes parameters allow 
for correlations between the horizontal and vertical polarizations of 
light in radiative transfer theory.  In the realistic high-redshift 
Universe, we do not expect correlations between states with $F=0$ and 
states with $F=1$, because their energy splitting implies any such 
correlations are destroyed on a timescale of $h/\Delta E\sim 10^{-9}\,$s 
much shorter than the collisional or radiative transition timescale.  
There may however be correlations between states with $F=1$ but different 
$M_F$, since these are degenerate (except for small effects such as Zeeman 
splitting; \citealt{2005MNRAS.359L..47C}).

This general problem can be simplified considerably for the special case 
where (A) the spin and velocity relaxation times are fast compared to the 
diffusion or free-streaming time across scales where bulk velocity or 
temperature gradients are significant; (B) the spin and velocity 
relaxation times are fast compared to the expansion, rotation, and 
shearing timescales of the fluid, and the Hubble and Compton-heating 
timescales; (C) an isotropic radiation field with smooth frequency 
dependence (such as the CMB); and (D) collisional transitions are 
dominated by spin exchange mechanisms.  These are very good approximations 
in most of the Universe (i.e. regions of near mean density) during the 
dark ages.  In these regions, the spin relaxation time, considering 
radiative transitions alone, is $T_\star/4AT_\gamma=2[40/(1+z)]\,$kyr, 
which is much faster than the Hubble time, 100$[40/(1+z)]^{3/2}\,$Myr 
(note that collisions at high redshift act to decrease the spin relaxation 
time.)  The velocity relaxation time, i.e. the timescale for the H atoms 
to relax to a Maxwellian distribution, is $120\,$kyr at 
$1+z=40$.\footnote{The number given here is the inverse decay constant of 
the slowest-decaying perturbation from a Maxwellian distribution. 
Technically, it is computed as the inverse of the smallest positive 
eigenvalue of $X'_{\Sigma\Sigma}$ defined in Eq.~(\ref{eq:x-prime}).} In 
regions of near mean density, the expansion time is of order the Hubble 
time, and the rotation and shearing timescales are longer; thus condition 
(B) should be valid.  Prior to collapse the expectation is that the 
velocity and temperature are smooth on scales smaller than the Jeans 
length; since the Jeans length is roughly the distance an atom can travel 
in a Hubble time (neglecting collisions), it follows that if (B) is valid, 
then (A) should be valid in either the diffusion or free-streaming cases.  
The CMB is isotropic and has blackbody frequency dependence, so (C) should 
be valid unless the 21-cm radiation adds (or subtracts) significantly to 
(or from) the CMB temperature.  The mean 21-cm brightness temperature from 
high-redshift gas has been computed to be $<1$ per cent of the CMB 
temperature (e.g. \citealt{2004PhRvL..92u1301L}), so (C) should be valid 
in regions of order mean density.  Finally, the ratio of dipolar 
relaxation to spin exchange rate is of order $10^{-2}$ or less at all 
relevant temperatures \citep{2005ApJ...622.1356Z}, so (D) is valid.

The simplifications allowed by (A--D) are as follows.  Condition (A) 
implies that the 
problem of solving for the density matrix and associated master equation 
can be solved locally, eliminating the position $\br$ from the problem.  
Condition (B) implies that we can treat the determination of the 
distribution function within the steady-state approximation, rather than 
evolving the Boltzmann equation from the recombination epoch forward to 
the epoch of interest.
Condition (C) implies that the radiative transition rates are independent 
of the magnitude and direction of the atom's velocity and are isotropic 
(i.e. no value of $M_F$ is favoured); the smoothness of the frequency 
dependence is important because otherwise it would be possible for atoms 
moving in different directions to have different 21-cm excitation rates 
due to the Doppler shift.  Condition (D) further implies that there is no 
coupling between the direction of spins and the direction of velocities.  
Thus in addition to being isotropic under rotations of both velocities 
and spins together, the problem is isotropic under rotations of velocities 
and spins separately.  Therefore once a steady-state solution is reached, 
one can assume that the density matrix $\rho_{FM_F,F'M'_F}(\bv)$ is 
isotropic under rotations of the spins, i.e. it takes the form
\beq
\rho_{FM_F,F'M'_F}(\bv) = \left(\begin{array}{cccc}
f_0(\bv) & 0 & 0 & 0 \\
0 & f_1(\bv)/3 & 0 & 0 \\
0 & 0 & f_1(\bv)/3 & 0 \\
0 & 0 & 0 & f_1(\bv)/3
\end{array}\right)
\eeq
in the $\{|00\rangle,|1-1\rangle,|10\rangle,|11\rangle\}$ basis.  Here 
$f_F(\bv)$ is the level-resolved distribution function, and it represents 
the number of hydrogen atoms in the hyperfine level 
with total spin $F\in\{0,1\}$ per unit volume in physical space, per unit 
volume in velocity space.  The total number density of hydrogen atoms is 
simply
\beq
\nHI = \sum_{F=0}^1 \int f_F(\bv)\,\rmd^3\bv.
\eeq

The key problem solved in this paper is to determine the distribution 
function $f_F(v)$ under the above-described conditions (A--D) by 
constructing the Boltzmann equation and finding its steady-state solution, 
and the associated 21-cm emission.  Computationally, this is a significant 
simplification as compared to solving for $\rho_{FM_F,F'M'_F}(\br, \bv)$.  
The limitations of this approach should be kept in mind however.  The most 
serious limitation is condition (C), which can be violated if the 21-cm 
line radiation ever becomes comparable in intensity to the CMB (i.e. if 
the line brightness temperature becomes $\sim T_\gamma$).  This does not 
occur in regions of the Universe near mean density, but it may happen in 
minihaloes (e.g. \citealt{2005astro.ph.12516S}) and thus our results 
should be used with caution in this case.

\subsection{Evolution equations}

The distribution function evolves according to several effects including 
radiative transitions, collisions with various species, Lyman-$\alpha$ 
scattering, and bulk velocity gradients.  In 
general one can decompose the rate of change of the distribution function 
into these effects as
\beq
\dot f_F(\bv) = \dot f_F^{(\rad)}(\bv)
 + \dot f_F^{(\col,\rmH-\rmH)}(\bv)
 + \dot f_F^{(\col,\rmH-\rmHe)}(\bv)
 + \dot f_F^{(\col,\rmH-e^-)}(\bv)
 + \dot f_F^{(\col,\rmH-\rmH^+)}(\bv)
 + \dot f_F^{(\lya)}(\bv)
 + \dot f_F^{(\bv_b)}(\bv).
\label{eq:boltz}
\eeq
The term associated with the bulk baryonic fluid velocity $\bv_b$ is,
\beq
\dot f_F^{(\bv_b)}(\bv) = a^{-1} v^i \frac{\partial(v_b)^j}{\partial r^i}
  \frac{\partial f_F(\bv)}{\partial v^j},
\eeq
because the velocity $\bv$ of the atom relative to the baryonic fluid 
changes as the atom moves to a region of different $\bv_b$ even if there 
are no forces acting on the atom.  (The Hubble expansion is considered 
to be part of the velocity gradient.)  We neglect this term since the 
associated timescale is of order the Hubble timescale (in mean-density 
regions) or the collapse timescale (in collapsing regions), either of 
which is long compared to the spin-relaxation timescale.

Of the remaining terms, the radiative transition term is the simplest to 
compute.  The 
photon phase space density in the 21-cm line is given by the blackbody 
formula
${\cal N} = (\rme^{T_\star/T_\gamma}-1)^{-1}$.
We neglect the change in momentum of the atom due to absorption or 
emission of a 21-cm photon, which is an excellent approximation
since the recoil energy in temperature units is 
$(h\nu_{10}/c)^2/2k_Bm_\rmH = 2\times 10^{-16}\,$K.
Then the radiative term becomes
\beq
\dot f_1^{(\rad)}(\bv) = -\dot f_0^{(\rad)}(\bv)
 = -A(1+{\cal N})f_1(\bv) + 3A{\cal N} f_0(\bv).
\label{eq:dfr}
\eeq

The collisional term is more complicated.  For the case of H-H collisions, 
we have
\beqa
\dot f_F^{(\col,\rmH-\rmH)}(\bv) &=& - \sum_{F'=0}^1 \int
  K_{FF'}(\bv'-\bv) f_F(\bv) f_{F'}(\bv') \,\rmd^3\bv'
\nonumber \\
&& + \frac{1}{2}
\sum_{F',F''} \int K_{F'F''}(\bv''-\bv') f_{F'}(\bv')f_{F''}(\bv'')
  p_{F|F'F''}(\bv|\bv',\bv'') \rmd^3\bv'\,\rmd^3\bv''.
\label{eq:dfc}
\eeqa
Here $K_{F'F''}$ is the product of cross section and relative velocity for 
two hydrogen atoms in the $F'$ and $F''$ level, and 
$p_{F|F'F''}(\bv|\bv',\bv'')$ is the probability density for such a 
collision to produce a final-state H atom in the $F$ level with velocity 
$\bv$.  The factor of $1/2$ in the second term avoids double-counting of 
collisions since the integral over $\rmd^3\bv'\,\rmd^3\bv''$ extends over 
all of ${\mathbb R}^6$.  The probability $p_{F|F'F''}(\bv|\bv',\bv'')$ 
must integrate to 2 since the final state has two H atoms:
\beq
\sum_{F=0}^1 \int p_{F|F'F''}(\bv|\bv',\bv'') \,\rmd^3\bv = 2.
\label{eq:final2}
\eeq
A similar but simpler collision term applies for H-He, H-H$^+$, and 
H-$e^-$ collisions, but of course in these cases Eq.~(\ref{eq:final2}) 
integrates to 1, and there is no summation over hyperfine levels of the 
target (i.e. the $F'$ summation in Eq.~\ref{eq:dfc} is not needed).  
Since He is a spin singlet, spin exchange does not occur in H-He 
collisions, however in general $\dot f_F^{(\col,\rmH-\rmHe)}(\bv)\neq0$ 
because the collision changes the velocity of the H atom.  For this reason 
we have included helium, but its effects turn out to be small ($<0.1$ per 
cent in the line FWHM and total emissivity).  Note that while the 
direct effect of He collisions is small the 21-cm emissivity 
still depends on the He fraction $Y_{\rm He}$ because it sets the density 
of H atoms and  the evolution of $T_k$.

In this paper we will neglect the atom-electron and atom-proton collision 
terms as we are interested in the period before UV or X-ray sources become 
important and begin to produce ionization.  There are some free electrons 
and protons (abundance $x_e\approx 2\times 10^{-4}$; 
\citealt{1999ApJ...523L...1S}) during the Dark Ages since the cosmological 
recombination does not run to completion.  A rough estimate of the 
importance of the electrons can be obtained by the previously computed 
thermal-averaged spin-change ($F=1\rightarrow 0$) cross sections $\langle 
\sigma_{10}v\rangle$; using the values of the H-H cross section in 
\citet{2005ApJ...622.1356Z} and the H-$e^-$ cross section of 
\citet{1966P&SS...14..929S}, we find $x_e\langle 
\sigma_{10}v\rangle_{\rmH-e^-}/\langle 
\sigma_{10}v\rangle_{\rmH-\rmH}\simeq0.01$ at $T_k=10\,$K ($z=21$) and 
less at higher redshifts.  We thus conclude that electrons change the 
collisional spin-change rate at the sub per cent level during the epoch of 
interest.  The spin-change rate due to collisions with protons is even 
smaller \citep{1966P&SS...14..929S}.

\subsection{Perturbation theory}

In general the evolution equations are nonlinear and solving them is 
highly nontrivial.  However, in the cases of interest, $T_\star=68.2\,$mK 
is much less than either $T_k$ or $T_\gamma$ and the equations can be 
linearized by expanding around the thermal equilibrium solution.  We
define
\beq
f_F^{(\rth)}(\bv) = \nHI y_F(T_k) \Phi_{T_k}(\bv).
\label{eq:thermal}
\eeq
to be the thermal distribution function at temperature 
$T_k$.\footnote{Note that $f_F^{(\rth)}(\bv)$ is computed with the spins 
thermalized at $T_k$.} We can now define the perturbation
\beq
\label{eq:xipertf}
\xi_F(\bv) = f_F(\bv) - f_F^{(\rth)}(\bv).
\eeq
Since $f_F^{(\rth)}(\bv)$ is a constant, all of the evolution equations 
can be trivially expressed in terms of $\xi_F(\bv)$ and its time 
derivative.   In terms of these new variables the radiative transition 
term becomes
\beq
\dot\xi_1^{(\rad)}(\bv)
= A\frac{-3\rme^{-T_\star/T_k}(1+{\cal N})
  + 3{\cal N}}
{3\rme^{-T_\star/T_k}+1} \nHI\Phi_{T_k}(\bv)
  - A(1+{\cal N})\xi_1(\bv) + 3A{\cal N}\xi_0(\bv),
\eeq
and working to lowest order in $T_\star$ this simplifies to
\beqa
\dot\xi_1^{(\rad)}(\bv) =
\frac{3}{4}A\left(\frac{T_\gamma}{T_k}-1\right)
\nHI  \Phi_{T_k}(\bv)
  - \frac{T_\gamma}{T_\star}A[\xi_1(\bv)-3\xi_0(\bv)].
\label{eq:dxr}
\eeqa
The equation for the $F=0$ levels is related to this by 
$\dot\xi_0^{(\rad)}(\bv)=-\dot\xi_1^{(\rad)}(\bv)$.

The H-H collisional term in the evolution equation is evaluated by 
substituting Eq.~(\ref{eq:xipertf}) into Eq.~(\ref{eq:dfc}) and expanding 
to leading order in $\xi$.  In principle, the collision term includes 
zeroth, first, and second-order terms.  However, the zeroth-order terms 
vanish because if the velocities and spin level populations were thermal 
then the distribution function would be independent of time.  The 
remaining terms are
\beqa
\dot \xi_F^{(\col,\rmH-\rmH)}(\bv) \!\!\! &=&
   - \sum_{F'=0}^1 \int
  K_{FF'}(\bv'-\bv) \xi_F(\bv) f_{F'}^{(\rth)}(\bv') \,\rmd^3\bv'
   - \sum_{F'=0}^1 \int
  K_{FF'}(\bv'-\bv) f^{(\rth)}_F(\bv) \xi_{F'}(\bv') \,\rmd^3\bv'
\nonumber \\
&& + \sum_{F',F''} \int K_{F'F''}(\bv''-\bv') 
\xi_{F'}(\bv')f^{(\rth)}_{F''}(\bv'')
  p_{F|F'F''}(\bv|\bv',\bv'') \rmd^3\bv'\,\rmd^3\bv''.
\nonumber \\
&& + \sum_{F',F''} \int K_{F'F''}(\bv''-\bv') 
\xi_{F'}(\bv')\xi_{F''}(\bv'')
  p_{F|F'F''}(\bv|\bv',\bv'') \rmd^3\bv'\,\rmd^3\bv''.
\label{eq:dxc}
\eeqa
So long as the kinetic and radiation temperatures are $\gg T_\star$, we 
will have $|\xi_F(\bv)|\ll f^{(\rth)}_F(\bv)$ and the 
second-order term on the last line of this equation can be neglected.

\section{Basis function method}
\label{sec:basis}

It is convenient at this point to expand the perturbation variable in a 
basis set.  We write
\beq
\xi_F(\bv) = \sum_n \xi_{Fn} \phi_n(\bv),
\label{eq:xif}
\eeq
where $\phi_n(\bv)$ form a set of basis functions satisfying the 
orthonormality relation,
\beq
\int \phi_n^\ast(\bv)\phi_{n'}(\bv)\,\rmd^3\bv = 
\left( \frac{\mH}{4\pi k_BT_k} \right)^{3/2}
\delta_{nn'}.
\label{eq:ortho}
\eeq
The normalization has been chosen so that the thermal distribution is one 
of the basis modes: $\phi_0(\bv) = \Phi_{T_k}(\bv)$.  The remaining modes 
are chosen to be the Gaussian-weighted Hermite polynomials,
\beq
\phi_n(\bv) = 2^{-n-5/2}\pi^{-3/2}[(2n+1)!]^{-1/2}
\frac{1}{\sigma^2v} 
H_{2n+1}\left( \frac v\sigma \right)
\rme^{-v^2/2\sigma^2},
\label{eq:phi-n}
\eeq
which indeed satisfy Eq.~(\ref{eq:ortho}). The 
$\{\phi_n\}_{n=0}^\infty$ are complete only for spherically symmetric 
functions (functions that depend only on the magnitude $v$) and in 
general additional basis functions 
with angular dependence must be included. However, since our problem is 
spherically symmetric, there is 
no need to include them.

Written in terms of basis functions Eq.~(\ref{eq:dxr}) becomes
\beq
\dot\xi_{1n}^{(\rad)} = 
\frac{3}{4}A\left(\frac{T_\gamma}{T_k}-1\right)\nHI\delta_{n,0}
 - \frac{T_\gamma}{T_\star} A(\xi_{1n}-3\xi_{0n}),
\label{eq:rads1}
\eeq
and $\dot\xi_{0n}^{(\rad)} = -\dot\xi_{1n}^{(\rad)} $.
Evaluation of the collisional rates is aided by defining several 
integrals.  We define for the H-H collisions,
\beqa
X^{(\col,\rmH-\rmH)}_{Fn,F'n'} &=& ( 4\pi\sigma^2)^{3/2}\Biggl[ 
\int \sum_{F''} y_{F''}(T_k)
K_{F'F''}(\bv''-\bv)
  \phi_n(\bv)\phi_{n'}(\bv)\phi_0(\bv'')
  \rmd^3\bv\,\rmd^3\bv''
\nonumber \\ && + \int y_{F}(T_k) K_{F'F}(\bv-\bv')
  \phi_n(\bv)\phi_{n'}(\bv')\phi_0(\bv)
  \rmd^3\bv\,\rmd^3\bv'
\nonumber \\ && -\int \sum_{F''} y_{F''}(T_k)
K_{F'F''}(\bv''-\bv') p_{F|F'F''}(\bv|\bv',\bv'')
  \phi_n(\bv)\phi_{n'}(\bv')\phi_0(\bv'')
  \rmd^3\bv\,\rmd^3\bv'\,\rmd^3\bv'' \Biggr].
\label{eq:x-h}
\eeqa
The first-order terms in Eq.~(\ref{eq:dxc}) then can be written compactly as
\beq
\dot\xi_{Fn}^{(\col,\rmH-\rmH)} = - \nHI\sum_{F'} 
X^{(\col,\rmH-\rmH)}_{Fn,F'n'}\xi_{F'n'}.
\eeq

Before proceeding, we note that the Maxwellian velocity distribution of
Eq.~(\ref{eq:maxwellian}) assumed in the standard calculation corresponds 
to a perturbation
\beq
\xi_F(\bv)\Tsapprox = [y_F(T_s)-y_F(T_k)]\nHI\Phi_{T_k}(\bv)
= \frac{3T_\star}{16}(-1)^F(T_s^{-1}-T_k^{-1})\nHI\Phi_{T_k}(\bv),
\eeq
or equivalently
\beq
\xi_{Fn}\Tsapprox = \frac{3T_\star}{16}(-1)^F(T_s^{-1}-T_k^{-1})\nHI\delta_{n,0}.
\label{eq:thermal-xi}
\eeq
In this case only the lowest basis mode ($n=0$) is excited.  In the full 
calculation, we will find that in general $\xi_{Fn}$ can be nonzero for 
any value of $n$.

\section{Evaluation of H-H collision term}
\label{sec:hhcoll}

\subsection{The relation of $K$ and $p$ to the cross section}
\label{ss:dcs}

Our objective here is to compute $K_{F'F''}(\bv''-\bv')$ and
$p_{F|F'F''}(\bv|\bv',\bv'')$.  These are related to the differential 
cross section as follows.  Defining
\beq
\bu = \frac{\bv''+\bv'}{2} \quad{\rm and}\quad
\bw = \bv''-\bv',
\eeq
we find that
\beq
K_{F'F''}(\bv''-\bv') = w\sigma_{F'F''}(w).
\eeq
The probability distribution for scattered products is
\beq
p_{F|F'F''}(\bv|\bv',\bv'') \,\rmd^3\bv =
\frac{\rmd P_{F|F'F''}}{\rmd\hat\bOmega}\,\rmd^2\hat\bOmega,
\label{eq:p-omega}
\eeq
where $\rmd P_{F|F'F''}/\rmd\hat\bOmega$ is the joint probability 
distribution for the final outgoing angle $\hat\bOmega$ and spin $F$
(which in general depends on the angle between $\hat\bOmega$ and 
$\hat\bw$) and
\beq
\bv = \bu + \frac{1}{2}w \hat\bOmega.
\label{eq:vuw}
\eeq
The final relative velocity is $w$ within the approximation of elastic 
scattering (valid when $T_\star\ll T_k$) where we neglect the hyperfine energy defect in comparison 
with the kinetic energy.  Note once again that the joint probability distribution integrates to 2 
because there are 2 H atoms:
\beq
\sum_{F=0}^1
\int \frac{\rmd P_{F|F'F''}}{\rmd\hat\bOmega}\,\rmd^2\hat\bOmega
= 2.
\label{eq:h-norm}
\eeq
The explicit expressions for the differential cross section can be
derived by techniques described in e.g. \citet{1966P&SS...14..937S}
and are summarized in Appendix~\ref{app:cross} for reference.

\subsection{Evaluation of integrals}

We focus our attention on the third integral appearing in brackets in 
Eq.~(\ref{eq:x-h}):
\beq
{\cal I} = \int \sum_{F''} y_{F''}(T_k)
K_{F'F''}(\bv''-\bv') p_{F|F'F''}(\bv|\bv',\bv'')
  \phi_n(\bv)\phi_{n'}(\bv')\phi_0(\bv'')\,
  \rmd^3\bv\,\rmd^3\bv'\,\rmd^3\bv'' \, .
\eeq
If we can evaluate this integral the first two 
integrals can also be evaluated by replacing 
$p_{F|F'F''}(\bv|\bv',\bv'')$ with $\delta_{FF'}\delta^{(3)}(\bv-\bv')$ or 
$\delta_{FF''}\delta^{(3)}(\bv-\bv'')$ respectively.  In terms of the 
differential cross sections of \S\ref{ss:dcs} this corresponds to setting 
$\rmd P_{F|F'F''}/\rmd\hat\bOmega$ equal to 
$\delta_{FF'}\delta^{(2)}(\hat\bOmega+\hat\bw)$ or
$\delta_{FF''}\delta^{(2)}(\hat\bOmega-\hat\bw)$ respectively.

Upon substituting Eq.~(\ref{eq:p-omega}) into ${\cal I}$, we get
\beq
{\cal I} = \int \sum_{F''} y_{F''}(T_k)
K_{F'F''}(w) \frac{\rmd P_{F|F'F''}}{\rmd\hat\bOmega}
  \phi_n\left(\bu + \frac{1}{2}w \hat\bOmega\right)
  \phi_{n'}\left(\bu - \frac{1}{2}\bw \right)
  \phi_0\left(\bu + \frac{1}{2}\bw \right)\,
  \rmd^3\bu\,\rmd^3\bw\,\rmd^2\hat\bOmega.
\eeq
Since we are considering only the spherically symmetric modes 
$\phi_n$ this integral can be simplified by symmetry considerations.  
We first split the integration over $\bw$ into integrals over magnitude $w$ and 
direction $\hat\bw$: $\rmd^3\bw=w^2\,\rmd w\,\rmd\hat\bw$.  If we then evaluate the $\rmd^3\bu\,\rmd w\,\rmd^2\hat\bOmega$ integral the result 
must be independent of the direction of $\hat\bw$.  We can thus replace 
$\hat\bw$ with any unit vector, say the unit vector in the third 
coordinate direction $\be_3$.  This gives
\beq
{\cal I} = 4\pi\int \sum_{F''} y_{F''}(T_k)
K_{F'F''}(w) \frac{\rmd P_{F|F'F''}}{\rmd\hat\bOmega}
  \phi_n\left(\bu + \frac{1}{2}w \hat\bOmega\right)
  \phi_{n'}\left(\bu - \frac{1}{2}w\be_3 \right)
  \phi_0\left(\bu + \frac{1}{2}w\be_3 \right)\, w^2\,
  \rmd^3\bu\,\rmd w\,\rmd^2\hat\bOmega.
\label{eq:3}
\eeq
By a similar argument if we decompose $\hat\bOmega$ into spherical 
coordinates,
\beq
\bOmega = \cos\theta\,\be_3 + \sin\theta\,(\cos\varphi\,\be_1
  + \sin\varphi\,\be_2),
\label{eq:sphere}
\eeq
then $\varphi$ can be integrated out to yield
\beqa
{\cal I} &=& 8\pi^2\int \sum_{F''} y_{F''}(T_k)
K_{F'F''}(w) \frac{\rmd P_{F|F'F''}}{\rmd\hat\bOmega}
  \phi_n\left[\bu+\frac{w}{2}(\cos\theta\be_3+\sin\theta\be_1)\right]
  \phi_{n'}\left(\bu - \frac{w}{2}\be_3 \right)
  \phi_0\left(\bu + \frac{w}{2}\be_3 \right)
\nonumber \\ && \times
 w^2\sin\theta\,
  \rmd^3\bu\,\rmd w\,\rmd\theta.
\label{eq:i}
\eeqa

Equation~(\ref{eq:i}) contains a 5-dimensional integral over $\bu$, $w$, 
and $\theta$.  In order to solve it efficiently we note that the 
integrand consists of two parts: one is the differential cross section 
$K_{F'F''} \rmd P_{F|F'F''}/\rmd\hat\bOmega$, which does not depend on
$\bu$, $n$, or $n'$; and the other is the product 
of basis functions, which does not depend on $F$, $F'$, or $F''$ and can 
be evaluated independently of the atomic physics.  Therefore we define the 
integral of three basis functions
\beq
C_{nn'}(w,\theta) = \int 
  \phi_n\left[\bu+\frac{w}{2}(\cos\theta\be_3+\sin\theta\be_1)\right]
  \phi_{n'}\left(\bu - \frac{w}{2}\be_3 \right)
  \phi_0\left(\bu + \frac{w}{2}\be_3 \right)
\,\rmd^3\bu.
\label{eq:c}
\eeq
It is readily seen that in our choice of basis, Eq.~(\ref{eq:phi-n}), the 
integrand takes on a special form.  The basis function $\phi_n(\bv)$ is a 
Gaussian $\rme^{-v^2/2\sigma^2}$ times a polynomial of order $2n$ in the 
components $(v_1,v_2,v_3)$ of $\bv$.  Therefore the integrand in 
Eq.~(\ref{eq:c}) is equal to a polynomial of order $2(n+n')$ times the 
product of Gaussians
\beqa
&&
\exp\left\{ -\frac{1}{2\sigma^2} 
\left[\bu+\frac{w}{2}(\cos\theta\be_3+\sin\theta\be_1)\right]^2
\right\}
\exp \left[ -\frac{1}{2\sigma^2}
\left(\bu - \frac{w}{2}\be_3 \right)^2 \right]
\exp \left[ -\frac{1}{2\sigma^2}
\left(\bu + \frac{w}{2}\be_3 \right)^2 \right]
\nonumber \\ && \quad
\propto \exp \left\{-\frac{3}{2\sigma^2}
\left[
\bu + \frac{w}{6}(\cos\theta\be_3+\sin\theta\be_1)
\right]^2
\right\},
\eeqa
where the proportionality constant is independent of $\bu$.  In 1 
dimension the $N$-point Gauss-Hermite integration formula 
gives an exact result for products of polynomials of degree $<2N$ and 
Gaussians.  The same is true in 3 dimensions for the $N^3$-point 
integration formula obtained by performing Gauss-Hermite integration 
on each of the three coordinate axes.  In this case it is necessary to 
choose the weights corresponding to a centroid 
$w(\cos\theta\be_3+\sin\theta\be_1)/6$ 
and 1$\sigma$ width $\sigma/\sqrt{3}$, and take $N>n+n'$.  
The computational demand can be reduced by noting that
the integrand in Eq.~(\ref{eq:c}) is even under the substitution 
$u_2\rightarrow -u_2$.  Substituting in Eq.~(\ref{eq:c}) we obtain
\beq
{\cal I} = 8\pi^2 \int \sum_{F''} y_{F''}(T_k) K_{F'F''}(w)
\frac{\rmd P_{F|F'F''}}{\rmd\hat\bOmega}
C_{nn'}(w,\theta)\,w^2\sin\theta\,\rmd w\,\rmd \theta.
\eeq
As noted earlier, the remaining two integrals in Eq.~(\ref{eq:x-h}) can be 
evaluated by replacing $\rmd P_{F|F'F''}/\rmd\hat\bOmega$ with 
$\delta_{FF'}\delta^{(2)}(\hat\bOmega+\hat\bw)$ and 
$\delta_{FF''}\delta^{(2)}(\hat\bOmega-\hat\bw)$.
This yields
\beqa
X^{(\col,\rmH-\rmH)}_{Fn,F'n'} &=& 8\pi^2(4\pi\sigma^2)^{3/2}
\int \sum_{F''} y_{F''}(T_k) K_{F'F''}(w)
\left[
\delta_{FF'}\delta^{(2)}(\hat\bOmega+\hat\bw)
+\delta_{FF''}\delta^{(2)}(\hat\bOmega-\hat\bw)
-\frac{\rmd P_{F|F'F''}}{\rmd\hat\bOmega}
\right]
\nonumber \\ && \times
C_{nn'}(w,\theta)\,w^2\sin\theta\,\rmd w\,\rmd\theta.
\eeqa
The $\delta$ functions of $\hat\bOmega$ can be converted to functions of 
$\theta$ according to the standard prescription
\beq
2\pi \delta^{(2)}(\hat\bOmega+\hat\bw)\,\sin\theta
= \delta(\theta-\pi)
\quad{\rm and}\quad
2\pi \delta^{(2)}(\hat\bOmega-\hat\bw)\,\sin\theta
= \delta(\theta).
\label{eq:delta}
\eeq
Thus
\beqa
X^{(\col,\rmH-\rmH)}_{Fn,F'n'} &=&
4\pi \left( 4\pi\sigma^2 \right)^{3/2}
\Biggl[
\delta_{FF'}\int \sum_{F''} y_{F''}(T_k) K_{FF''}(w)
C_{nn'}(w,\pi)\,w^2\,dw
+\int y_{F}(T_k) K_{F'F}(w)
C_{nn'}(w,0)\,w^2\,dw
\nonumber \\ &&
-2\pi
\int \sum_{F''} y_{F''}(T_k) K_{F'F''}(w)
\frac{\rmd P_{F|F'F''}}{\rmd\hat\bOmega}
C_{nn'}(w,\theta)\,w^2\sin\theta\,\rmd w\,\rmd\theta\Biggr].
\eeqa
Since we are only computing $X^{(\col,\rmH-\rmH)}_{Fn,F'n'}$ to zeroth 
order 
in $T_\star$ it is permissible to set $y_F(T_k)\rightarrow 
(2F+1)/4$.  Making this substitution and recalling that 
$K_{F'F''}(w)=w\sigma_{F'F''}(w)$ we arrive at
\beqa
X^{(\col,\rmH-\rmH)}_{Fn,F'n'} &=&
\pi ( 4\pi\sigma^2)^{3/2}
\Biggl[\delta_{FF'}
\int \sum_{F''} (2F''+1) \sigma_{FF''}(w)
C_{nn'}(w,\pi)\,w^3\,\rmd w
+\int (2F+1) \sigma_{F'F}(w)
C_{nn'}(w,0)\,w^3\,\rmd w
\nonumber \\ &&
-2\pi
\int \sum_{F''} (2F''+1) \sigma_{F'F''}(w)
\frac{\rmd P_{F|F'F''}}{\rmd\hat\bOmega}
C_{nn'}(w,\theta)\,w^3\sin\theta\,\rmd w\,\rmd\theta\Biggr].
\label{eq:x-h2}
\eeqa

\section{Inclusion of helium}
\label{sec:he}

In reality the primordial baryonic matter did not contain only $^1$H but 
also $^4$He with an abundance $n(^4\rmHe)/n(^1\rmH)\approx 0.08$ as well 
as trace quantities ($<10^{-4}$) of $^2$H, $^3$He, and $^7$Li.  
Collisions with the rare species will be unimportant compared to 
collisions with $^1$H but for precision at the level of several percent it 
is worth considering the role of $^1$H--$^4$He collisions.  Helium is an 
electron spin singlet and hence it cannot undergo spin exchange in 
collisions with hydrogen; thus in the usual approximation of a single spin 
temperature it does not do anything.  If the spin-resolved velocity 
distribution is nonthermal however, collisions with helium can 
redistribute the velocities of the $F=0$ and $F=1$ hydrogen atoms and 
bring them closer to the Maxwellian distribution.  Indeed if the 
$^4$He/$^1$H ratio were very large, so that collisions with He atoms 
redistributed the velocities faster than collisions with H atoms could 
change the hyperfine level populations, then one would expect the 
spin-resolved H velocity distribution to be Maxwellian and the single spin 
temperature approximation to become exact.

In this section we expand the Boltzmann equation to include the effects of 
helium.  The first step is to introduce the Maxwellian distribution for 
helium and the appropriate basis modes for describing perturbations of the 
helium distribution function.  We may then write the appropriate terms in 
the Boltzmann equation to account for H-He and He-He collisions.

In analogy with Eq.~(\ref{eq:phi-n}) we define helium basis modes
\beq
\phi_{\rmHe,n}(\bv) = 
2^{-n-5/2}\pi^{-3/2}[(2n+1)!]^{-1/2}
\frac{\mHe}{k_BT_kv} H_{2n+1}\left( \frac{v}{\sqrt{k_BT_k/\mHe}} \right)
\rme^{-\mHe v^2/2k_BT_k},
\eeq
which satisfy the orthonormality relation
\beq
\int \phi_{\rmHe,n}^\ast(\bv)\phi_{\rmHe,n'}(\bv)\,\rmd^3\bv =
\left( \frac{\mHe}{4\pi k_BT_k} \right)^{3/2}
\delta_{nn'}.
\eeq
Note that the Maxwellian distribution is simply 
$\Phi_\rmHe(\bv)=\phi_{\rmHe,0}(\bv)$, and the thermal helium distribution 
function is $f_\rmHe^{(\rth)}(\bv)=\nHeI\Phi_\rmHe(\bv)$.  We may then 
define the perturbation
\beq
\xi_\rmHe(\bv) = f_\rmHe(\bv) - f_\rmHe^{(\rth)}(\bv)
  = \sum_n \xi_{\rmHe,n}\phi_{\rmHe,n}(\bv).
\eeq
We now seek a linearized form of the Boltzmann equation for the H-He and 
He-He terms in analogy to Eq.~(\ref{eq:x-h2}).  The spin exchange 
mechanism does not operate in H-He collisions since He is a spin singlet
($S=0$).  Therefore the H atom has the same value of $F$ before and after 
the collision, and the H-He collision term in the Boltzmann equation for H 
is
\beq
\dot f_F^{(\col,\rmH-\rmHe)}(\bv) = -\int K^{(\rmH-\rmHe)}(\bv'-\bv)
  f_F(\bv) f_{\rmHe}(\bv')\,\rmd^3\bv'
+ \int K^{(\rmH-\rmHe)}(\bv''-\bv') f_\rmHe(\bv')f_F(\bv'')
p_\rmH^{(\rmH-\rmHe)}(\bv|\bv',\bv'') \,\rmd^3\bv'\,\rmd^3\bv''.
\eeq
Here $K^{(\rmH-\rmHe)}(w)$ is the product of relative velocity and cross 
section at relative velocity $w$, and 
$p_\rmH^{(\rmH-\rmHe)}(\bv|\bv',\bv'')$ is the probability density for the 
final velocity of the H atom.  Writing this in terms of perturbation 
variables gives
\beqa
\dot\xi_F^{(\col,\rmH-\rmHe)}(\bv) &=&
-\int K^{(\rmH-\rmHe)}(\bv'-\bv)
  [\xi_F(\bv) f^{(\rth)}_{\rmHe}(\bv')
  + f^{(\rth)}_F(\bv) \xi_{\rmHe}(\bv')]
\,\rmd^3\bv'
\nonumber \\ &&
+ \int K^{(\rmH-\rmHe)}(\bv''-\bv')
[f^{(\rth)}_\rmHe(\bv')\xi_F(\bv'')
+\xi_\rmHe(\bv') f^{(\rth)}_F(\bv'')]
p_\rmH^{(\rmH-\rmHe)}(\bv|\bv',\bv'') \,\rmd^3\bv'\,\rmd^3\bv'',
\label{eq:temphe1}
\eeqa
where we have used the fact that the zeroth-order terms on the right-hand 
side vanish.  We have also dropped the second-order terms in $\xi$.

Integration of Eq.~(\ref{eq:temphe1}) against $\phi_n(\bv)$ gives 
\beq
\dot\xi^{(\col,\rmH-\rmHe)}_{Fn} = -\nHeI \sum_{n'} Y_{nn'}
\xi^{(\col,\rmH-\rmHe)}_{Fn'}
 - \nHI \sum_{n'} 
X^{(\col,\rmH-\rmHe)}_{Fn,\rmHe\ n'}\xi^{(\col,\rmH-\rmHe)}_{\rmHe,n'},
\eeq
where the coefficients are:
\beqa
Y_{nn'} &=&
(4\pi\sigma^2)^{3/2}\Biggl[
\int K^{(\rmH-\rmHe)}(\bv'-\bv)
  \phi_n(\bv)\phi_{n'}(\bv)\phi_{\rmHe,0}(\bv')
\,\rmd^3\bv\,\rmd^3\bv'
\nonumber \\ &&
- \int K^{(\rmH-\rmHe)}(\bv''-\bv')
\phi_n(\bv)\phi_{\rmHe,0}(\bv')\phi_{n'}(\bv'')
p_\rmH^{(\rmH-\rmHe)}(\bv|\bv',\bv'') \,\rmd^3\bv
\,\rmd^3\bv'\,\rmd^3\bv''\Biggr]
\label{eq:y}
\eeqa
and
\beqa
X^{(\col,\rmH-\rmHe)}_{Fn,\rmHe\ n'}
&=& y_F(T_k)
(4\pi\sigma^2)^{3/2}\Biggl[
\int K^{(\rmH-\rmHe)}(\bv'-\bv)
  \phi_n(\bv)\phi_0(\bv)\phi_{\rmHe,n'}(\bv')
\,\rmd^3\bv\,\rmd^3\bv'
\nonumber \\ &&
- \int K^{(\rmH-\rmHe)}(\bv''-\bv')
\phi_n(\bv)\phi_{\rmHe,n'}(\bv')\phi_{0}(\bv'')
p_\rmH^{(\rmH-\rmHe)}(\bv|\bv',\bv'') \,\rmd^3\bv
\,\rmd^3\bv'\,\rmd^3\bv''\Biggr].
\eeqa
It will turn out that we need only explicitly evaluate $Y_{nn'}$ as 
$X^{(\col,\rmH-\rmHe)}_{Fn,\rmHe\ n'}$ will not affect the hydrogen level 
populations at linear order in perturbation theory because, as we show below,
in the steady state $\xi_{\rmHe\ n}=0$.

As in our analysis of H-H collisions the most efficient way to evaluate 
$Y_{nn'}$ is to focus on the second integral in brackets in 
Eq.~(\ref{eq:y}).  Substitution of 
$p_\rmH^{(\col,\rmH-\rmHe)}(\bv|\bv',\bv'')\rightarrow\delta^{(3)}(\bv-\bv'')$ 
recovers the first term.  The evaluation of the integral is most easily 
accomplished by considering the centre-of-mass and relative velocities 
rather than $\bv'$ and $\bv''$ as variables:
\beq
\bu = \frac{\mH\bv''+\mHe\bv'}{\mH+\mHe}
\quad{\rm and}\quad
\bw = \bv''-\bv';
\quad
\rmd^3\bv'\,\rmd^3\bv'' = \rmd^3\bu\,\rmd^3\bw.
\eeq
The probability distribution for the final velocity $\bv$ in the 
scattering event is given by the analogue of Eq.~(\ref{eq:p-omega})
\beq
p_\rmH^{(\col,\rmH-\rmHe)}(\bv|\bv',\bv'')\,\rmd^3\bv
= \frac{\rmd P}{\rmd\hat\bOmega}\rmd^2\hat\bOmega,
\eeq
and the specific relation between $\bv$ and $\bOmega$ is now
\beq
\bv = \bu + \frac{\mHe}{\mH+\mHe}\bw,
\eeq
which is different from Eq.~(\ref{eq:vuw}) since the He and H masses are 
unequal.  The second integral in Eq.~(\ref{eq:y}) then becomes
\beqa
{\cal I}_2&\equiv&
\int K^{(\rmH-\rmHe)}(\bv''-\bv')
\phi_n(\bv)\phi_{\rmHe,0}(\bv')\phi_{n'}(\bv'')
p_\rmH^{(\rmH-\rmHe)}(\bv|\bv',\bv'') \,\rmd^3\bv
\,\rmd^3\bv'\,\rmd^3\bv''
\nonumber\\&=&
\int K^{(\rmH-\rmHe)}(w)
\phi_n\left( \bu + \frac{\mHe}{\mH+\mHe}\hat\bOmega \right)
\phi_{\rmHe,0}\left(\bu - \frac{\mH}{\mH+\mHe}\bw\right)
\phi_{n'}\left(\bu + \frac{\mHe}{\mH+\mHe}\bw\right)
\frac{\rmd P}{\rmd\hat\bOmega}
\rmd^3\bu\,\rmd^3\bw\,\rmd^2\hat\bOmega
\nonumber\\&=&
4\pi\int K^{(\rmH-\rmHe)}(w)
\phi_n\left( \bu + \frac{\mHe w\hat\bOmega}{\mH+\mHe} \right)
\phi_{\rmHe,0}\left(\bu - \frac{\mH w\be_3}{\mH+\mHe}\right)
\phi_{n'}\left(\bu + \frac{\mHe w\be_3}{\mH+\mHe}\right)
\frac{\rmd P}{\rmd\hat\bOmega}
w^2\,\rmd^3\bu\,\rmd w\,\rmd^2\hat\bOmega.
\eeqa
In the last line we have used spherical symmetry to choose $\bw$ to 
be in the third coordinate direction and replace $\rmd^3\bw$ 
with $4\pi w^2\,\rmd w$.  If we write 
$\hat\bOmega$ in spherical coordinates (Eq.~\ref{eq:sphere}) and integrate 
out the $\varphi$ direction we then get
\beqa
{\cal I}_2 &=&
8\pi^2\int K^{(\rmH-\rmHe)}(w)
\phi_n\left[ \bu + \frac{\mHe w(\cos\theta\,\be_3
+\sin\theta\,\be_1)}{\mH+\mHe} \right]
\phi_{\rmHe,0}\left(\bu - \frac{\mH w\be_3}{\mH+\mHe}\right)
\phi_{n'}\left(\bu + \frac{\mHe w\be_3}{\mH+\mHe}\right)
\nonumber \\ && \times
\frac{\rmd P}{\rmd\hat\bOmega}
w^2\sin\theta\,\rmd^3\bu\,\rmd w\,\rmd\theta.
\eeqa
Once again it is possible to separate the purely kinematic terms in this 
integral from the cross sections.  We define
\beq
D_{nn'}(w,\theta) = \int
\phi_n\left[ \bu + \frac{\mHe w(\cos\theta\,\be_3
+\sin\theta\,\be_1)}{\mH+\mHe} \right]
\phi_{\rmHe,0}\left(\bu - \frac{\mH w\be_3}{\mH+\mHe}\right)
\phi_{n'}\left(\bu + \frac{\mHe w\be_3}{\mH+\mHe}\right)
\,\rmd^3\bu,
\eeq
so that
\beq
{\cal I}_2 = 8\pi^2\int K^{(\rmH-\rmHe)}(w)
\frac{\rmd P}{\rmd\hat\bOmega}
D_{nn'}(w,\theta)
\,w^2\sin\theta\,\rmd w\,\rmd\theta.
\eeq
Just as for $C_{nn'}$, it is possible to evaluate $D_{nn'}$ exactly by 
Gauss-Hermite integration on each of the three coordinate axes.  However 
in this case the centroid of integration is at 
$-\mH\mHe w(\cos\theta\,\be_3+\sin\theta\,\be_1)/[(\mH+\mHe)(2\mH+\mHe)]$ 
and the $1\sigma$ width is $\sqrt{k_BT_k/(2\mH+\mHe)}$.

With this value for ${\cal I}_2$, and recalling that the first integral in 
Eq.~(\ref{eq:y}) can be obtained from the second by the replacement 
$p_\rmH^{(\col,\rmH-\rmHe)}(\bv|\bv',\bv'')
\rightarrow\delta^{(3)}(\bv-\bv'')$, 
or equivalently $\rmd 
P/\rmd\hat\bOmega\rightarrow\delta^{(2)}(\hat\bOmega-\hat\bw)$, we find
\beqa
Y_{nn'} &=&
(4\pi\sigma^2)^{3/2}\Biggl[
8\pi^2\int K^{(\rmH-\rmHe)}(w)
\delta^{(2)}(\hat\bOmega-\hat\bw)
D_{nn'}(w,\theta)
\,w^2\sin\theta\,\rmd w\,\rmd\theta
\nonumber \\ &&
-8\pi^2\int K^{(\rmH-\rmHe)}(w)
\frac{\rmd P}{\rmd\hat\bOmega}
D_{nn'}(w,\theta)
\,w^2\sin\theta\,\rmd w\,\rmd\theta
\Biggr].
\eeqa
Using Eq.~(\ref{eq:delta}) this simplifies further to
\beq
Y_{nn'} =
(4\pi\sigma^2)^{3/2}\left[
\int K^{(\rmH-\rmHe)}(w)
D_{nn'}(w,0)
\,w^2\rmd w
-2\pi\int K^{(\rmH-\rmHe)}(w)
\frac{\rmd P}{\rmd\hat\bOmega}
D_{nn'}(w,\theta)
\,w^2\sin\theta\,\rmd w\,\rmd\theta
\right] \, ,
\label{eq:y-compute}
\eeq
which is now suitable for numerical evaluation.

\section{Solution of the Boltzmann equation}
\label{sec:solveboltz}

We are now interested in the steady-state solution to 
Eq.~(\ref{eq:boltz}).  For simplicity we restrict our attention to 
the case of a purely atomic gas of H and He and no \lya\ radiation.  
These conditions are expected to pertain to the era before the first 
sources of ultraviolet and X-ray radiation.

\subsection{The solution}

As currently constructed the Boltzmann equation can be written as
\beq
\dot\bxi = -\mathbfss X\bxi + \bmath S,
\label{eq:mat_boltzmann}
\eeq
where $\mathbfss X$ is the relaxation matrix and $\bmath S$ is the forcing 
term.  The dimension of the vectors $\bxi$ and $\bmath S$ is $3N$, where 
$N$ is the number of basis modes used. There are $N$ modes each for 
the perturbations in the hydrogen $F=0$ phase space density, for hydrogen 
$F=1$, and for helium.  The forcing term is purely radiative 
\beq
S_{1n} = -S_{0n} = 
\frac{3}{4}A\left(\frac{T_\gamma}{T_k}-1\right)\nHI\delta_{n,0},
\label{eq:rads2}
\eeq
and can be read off from Eq.~(\ref{eq:rads1}) along with $S_{\rmHe,n}=0$.  The relaxation matrix is
\beq
\mathbfss X = \mathbfss X^{(\rad)} + \nHI\mathbfss X^{(\col,\rmH-\rmH)}
  + \nHI\mathbfss X^{(\col,\rmH-\rmHe)} + \nHI\mathbfss
  X^{(\col,\rmHe-\rmHe)}.
\eeq
The various contributions to the relaxation matrix exhibit sparseness 
patterns when broken into the blocks corresponding to H$(F=0)$, H$(F=1)$, 
and He.  This turns out to be very useful in solving the Boltzmann 
equation because only some of the rates are relevant.  The radiative 
contribution can be read 
off from Eq.~(\ref{eq:rads1}) and it breaks down as
\beq
\mathbfss X^{(\rad)} = \frac{T_\gamma}{T_\star}A
\mtt{\mathbfss I}{-3\,\mathbfss I}{0}{-\mathbfss I}{3\,\mathbfss I}{0}
{0}{0}{0},
\label{eq:x1}
\eeq
where $\mathbfss I$ is the $N\times N$ identity matrix.  The H-H collision 
term only has non-zero entries for the hydrogen atoms,
\beq
\mathbfss X^{(\col,\rmH-\rmH)}
= \mtt{\mathbfss X^{(\col,\rmH-\rmH)}_{00}}
{\mathbfss X^{(\col,\rmH-\rmH)}_{01}}{0}
{\mathbfss X^{(\col,\rmH-\rmH)}_{10}}
{\mathbfss X^{(\col,\rmH-\rmH)}_{11}}{0}{0}{0}{0}.
\label{eq:x2}
\eeq
The H-He collision term has a particular structure due to the elastic 
cross section being the same for H$(F=0)$ and H$(F=1)$, and because the 
spin-exchange cross section vanishes:
\beq
\mathbfss X^{(\col,\rmH-\rmHe)}
= \mtt{(\nHeI/\nHI)\mathbfss Y}{0}
{\mathbfss X^{(\col,\rmH-\rmHe)}_{0\rmHe}}
{0}{(\nHeI/\nHI)\mathbfss Y}
{\mathbfss X^{(\col,\rmH-\rmHe)}_{1\rmHe}}
{\mathbfss X^{(\col,\rmH-\rmHe)}_{\rmHe0}}
{\mathbfss X^{(\col,\rmH-\rmHe)}_{\rmHe1}}{
\mathbfss X^{(\col,\rmH-\rmHe)}_{\rmHe\rmHe}},
\label{eq:x3}
\eeq
where ${\mathbfss X}^{(\col,\rmH-\rmHe)}_{\rmHe0} = 
{\mathbfss X}^{(\col,\rmH-\rmHe)}_{\rmHe1}$, 
and ${\mathbfss Y}$ is the matrix of H-He collision integrals defined in 
Eq.~(\ref{eq:y}).

Finally, for the He-He term, only the ${\mathbfss 
X}^{(\col,\rmHe-\rmHe)}_{\rmHe\rmHe}$ block is nonzero.

The key to solving the Boltzmann equation is to introduce the change of 
basis from the abundances of H$(F=0)$ and H$(F=1)$ to the new modes
\beq
\xi'_{\Delta,n} = \xi_{1,n}-3\xi_{0,n}, {\rm ~}
\xi'_{\Sigma,n} = \xi_{1,n}+\xi_{0,n},
\rm {~and~}
\xi'_{\rmHe,n} = \xi_{\rmHe,n}.
\eeq
The new basis $\{\xi'_{\Delta,n},\xi'_{\Sigma,n},\xi'_{\rmHe,n}\}$ 
is related to the old basis $\{\xi_{0,n},\xi_{1,n},\xi_{\rmHe,n}\}$ 
by a matrix,
\beq
\bxi'=\mathbfss R\bxi,\quad
\mathbfss R = \mtt{-3\mathbfss I}{\mathbfss I}{0}{\mathbfss 
I}{\mathbfss I}{0}{0}{0}{\mathbfss I}.
\eeq
The relaxation matrices $\mathbfss X$ can be expressed in the new basis 
via $\mathbfss X'=\mathbfss{RXR}^{-1}$, and the source is expressed as 
$\bmath S'=\mathbfss R\bmath S$.  It is at this point that a 
simplification occurs.  Although $\mathbfss X$ is in general non-sparse 
(cf. Eqs.~\ref{eq:x1}--\ref{eq:x3}), the first column of blocks of 
$\mathbfss X'$ read as 
follows:
\beqa
\mathbfss X'_{\Delta\Delta} &=&
4\frac{T_\gamma}{T_\star}A\mathbfss I +
\frac\nHI4[3\mathbfss X^{(\col,\rmH-\rmH)}_{00}
-3\mathbfss X^{(\col,\rmH-\rmH)}_{01}
-\mathbfss X^{(\col,\rmH-\rmH)}_{10}
+\mathbfss X^{(\col,\rmH-\rmH)}_{11}]
+ \nHeI\mathbfss Y,
\nonumber \\
\mathbfss X'_{\Sigma\Delta} &=& \frac\nHI4[
-\mathbfss X^{(\col,\rmH-\rmH)}_{00}
+\mathbfss X^{(\col,\rmH-\rmH)}_{01}
-\mathbfss X^{(\col,\rmH-\rmH)}_{10}
+\mathbfss X^{(\col,\rmH-\rmH)}_{11}]
= \frac\nHI2\sum_{F,F'} (-1)^{F'-1}
\mathbfss X^{(\col,\rmH-\rmH)}_{FF'}
,{\rm ~and} \nonumber \\
\mathbfss X'_{\rmHe\Delta} &=& 0.
\label{eq:xp1}
\eeqa
The latter expression may be evaluated by substituting Eq.~(\ref{eq:x-h2})
in for $X^{(\col,\rmH-\rmH)}_{FF'}$.  This gives
\beqa
X'_{\Sigma n, \Delta n'} \!\! &=&
\pi (4\pi\sigma^2)^{3/2}
\frac\nHI4\Biggl[
\int\sum_{F,F'} (-1)^{F'-1}
 (2F+1) \sigma_{F'F}(w)
C_{nn'}(w,0)\,w^3\,\rmd w
\nonumber \\ &&
-2\pi
\int \sum_{F,F',F''} (-1)^{F'-1}(2F''+1) \sigma_{F'F''}(w)
\frac{\rmd P_{F|F'F''}}{\rmd\hat\bOmega}
C_{nn'}(w,\theta)\,w^3\sin\theta\,\rmd w\,\rmd\theta\Biggr].
\label{eq:temp1}
\eeqa
(The first term in the square brackets in Eq.~\ref{eq:x-h2} does not 
contribute because it trivially cancels when summed over $F'$.)  Using 
Eq.~(\ref{eq:sigma-H}), one may verify directly that
\beq
\sum_{F,F',F''} (-1)^{F'-1}(2F''+1) \sigma_{F'F''}(w)
\frac{\rmd P_{F|F'F''}}{\rmd\hat\bOmega}=0
\label{eq:temp2}
\eeq
for all $w$ and $\theta$ and hence the last integral in 
Eq.~(\ref{eq:temp1}) vanishes.  By integrating Eq.~(\ref{eq:temp2}) over 
angles, we find
\beq
2 \sum_{F',F''}(-1)^{F'-1}(2F''+1) \sigma_{F'F''}(w)
 = \int
\sum_{F,F',F''} (-1)^{F'-1}(2F''+1) \sigma_{F'F''}(w)
\frac{\rmd P_{F|F'F''}}{\rmd\hat\bOmega}\,\rmd\hat\bOmega
= 0.
\label{eq:temp3}
\eeq
Noting that the total cross section is symmetrical, 
$\sigma_{F'F''}(w)=\sigma_{F''F'}(w)$, and that $F''$ is a dummy variable 
in the summation in Eq.~(\ref{eq:temp3}), we see that in fact the first 
integral in Eq.~(\ref{eq:temp1}) also vanishes, so that $X'_{\Sigma n, 
\Delta n'}=0$.  Therefore the matrix $\mathbfss X'$, and in particular its first column, reads
\beq
\mathbfss X' = \mtt{\mathbfss X'_{\Delta\Delta}}{\mathbfss X'_{\Delta\Sigma}}{\mathbfss X'_{\Delta\rmHe}}
{0}{\mathbfss X'_{\Sigma\Sigma}}{\mathbfss X'_{\Sigma\rmHe}}{0}{\mathbfss X'_{\rmHe\Sigma}}{\mathbfss X'_{\rmHe\rmHe}} \, .
\label{eq:x-prime}
\eeq
As we are about to 
see, we will only need to evaluate the first column of $\mathbfss X'$.  The source vector in the new basis 
is
\beq
S'_{\Delta,n} = 3A\left(\frac{T_\gamma}{T_k}-1\right)\nHI\delta_{n,0},
\quad S'_{\Sigma,n}=0,\quad S'_{\rmHe,n}=0.
\label{eq:s-prime}
\eeq
Thus only one entry of the vector $\bmath S'$ is nonzero.

The particular sparseness patterns of $\mathbfss X'$ and $\bmath S'$ are 
useful because they imply that the evolution of $\xi'_{\Sigma,n}$ and 
$\xi'_{\rmHe,n}$ does not depend on $\xi'_{\Delta,n}$ and has no sources.  
Therefore these components of $\xi'$ -- and more generally, the 
corresponding distribution functions $f_0(\bv)+f_1(\bv)$ and 
$f_\rmHe(\bv)$ -- will relax to thermal equilibrium.  We have chosen to 
expand our distribution functions around the thermal equilibrium solution 
at temperature $T_k$ and hence we will have 
\beq
\xi'_{\Sigma,n}=\xi'_{\rmHe,n}=0.
\eeq
This result allows us to simplify the equation for $\xi'_{\Delta,n}$ to 
yield $\dot\xi'_{\Delta,n}=-\sum_{n'}X'_{\Delta n,\Delta n'}\xi'_{\Delta,
n'}+S'_{\Delta,n}$.  If we consider the steady-state solution with 
$\dot\xi'_{\Delta,n}=0$ we conclude that
\beq
\bxi'_\Delta = (\mathbfss X'_{\!\Delta\Delta})^{-1}\bmath S'_\Delta.
\label{eq:bxid}
\eeq

Numerical evaluation of Eq.~(\ref{eq:bxid}) can be simplified by 
exploiting the result $\mathbfss X'_{\Sigma\Delta}=0$ and using 
Eq.~(\ref{eq:xp1}) to write
\beq
\mathbfss X'_{\Delta\Delta}
= \mathbfss X'_{\Delta\Delta}-\mathbfss X'_{\Sigma\Delta}
=4\frac{T_\gamma}{T_\star}A\mathbfss I +
\nHI[\mathbfss X^{(\col,\rmH-\rmH)}_{00}
-\mathbfss X^{(\col,\rmH-\rmH)}_{01}]+ \nHeI\mathbfss Y,
\eeq
so that
\beq
\xi'_{\Delta,n} = 3A\left(\frac{T_\gamma}{T_k}-1\right)\nHI\left(\left\{
4\frac{T_\gamma}{T_\star}A\mathbfss I +
\nHI[\mathbfss X^{(\col,\rmH-\rmH)}_{00}
-\mathbfss X^{(\col,\rmH-\rmH)}_{01}]+ \nHeI\mathbfss Y\right\}^{-1}
\right)_{n0},
\label{eq:xin}
\eeq
where the subscript $()_{n0}$ denotes the entry in the row corresponding 
to the $n$th basis mode and the column corresponding to the $0$th mode.

\subsection{A subtlety: CMB heating of the gas}

Before continuing, we discuss one subtlety of the steady-state solution to 
the Boltzmann equation, Eq.~(\ref{eq:xin}).  The method of solution 
depends on the observation that ${\mathbfss X}'_{\Sigma\Delta}=0$, which 
we proved for the elastic approximation cross sections.  It is natural to 
ask whether ${\mathbfss X}'_{\Sigma\Delta}=0$ would still be true if we 
dropped the elastic approximation and used the full cross sections 
instead.  While the answer to this questions is in fact no, and this 
drastically alters the steady-state solution to 
Eq.~(\ref{eq:mat_boltzmann}) if the full (inelastic) cross sections are 
used, the solution of Eq.~(\ref{eq:xin}) remains the physical solution to 
Eq.~(\ref{eq:boltz}).

In addition to being an approximation for the cross sections, the elastic 
approximation (in the simple form used here) treats the scattering as 
elastic, i.e. the total kinetic energy of the atoms is the same before and 
after collision.  For this reason, the elastic approximation does not 
include heating of the gas by the CMB in the 21-cm line (the process 
in which the CMB heats the spins, which then collisionally transfer energy 
to the kinetic degrees of freedom of the gas).  Of course, if this process 
is taken into account, the only possible steady-state solution to Eq.~(\ref{eq:mat_boltzmann}) is for both 
the spin and kinetic degrees of freedom to come to thermal equilibrium 
with the CMB.  Recall, however, that in the above derivation we dropped 
the explicitly time-dependent terms associated with the Hubble expansion 
because the Hubble time during this epoch is much longer than the 
radiative or collision times.  As we show below, the 21-cm heating time is even 
\emph{longer} than the Hubble time and thus, while in a 
nonexpanding universe the spin and kinetic degrees of freedom would have 
enough time to thermalize with the CMB, in our expanding Universe such heating 
is extremely inefficient.  For this reason the solution to the Boltzmann 
equations derived in the previous section by neglecting processes with 
timescales longer than the Hubble time is the physical solution.

A relatively simple argument can be used to estimate the timescale 
for heating in the 21-cm line.  In the spin temperature approximation, 
we know that the net rate of collisional de-excitation per H atom is
\beq
\Gamma_{\rm de-ex} = \nHI [ \langle\sigma_{10}v\rangle(T_k) y_0(T_s)
- \langle\sigma_{01}v\rangle(T_k) y_1(T_s) ]
\approx \frac34 \nHI \langle\sigma_{10}v\rangle(T_k)
T_\star(T_k^{-1}-T_s^{-1}),
\eeq
where $\langle\sigma_{10}v\rangle(T_k)$ is the thermally averaged cross 
section for de-excitation, and the approximation makes use of 
Eq.~(\ref{eq:34}) and the principle of detailed balance 
$\langle\sigma_{01}v\rangle(T_k) = 
3e^{-T_\star/T_k}\langle\sigma_{10}v\rangle(T_k)$.  The heating rate in 
the 
21-cm line is then $k_BT_\star\Gamma_{\rm de-ex}$ per atom.  Comparing 
this to the heat capacity for a monatomic gas, $C_v=3k_B(1+f_\rmHe)/2$ per 
H atom, implies that the heating rate is
\beq
\dot T_k|_{21\,\rm cm} =
\frac{k_BT_\star\Gamma_{\rm de-ex}}{C_v}
= \frac{\nHI\langle\sigma_{10}v\rangle(T_k)\;
T_\star^2(T_k^{-1}-T_s^{-1})}{2(1+f_\rmHe)},
\eeq
implying a timescale for heating of the gas
\beq
t_h \equiv \frac{T_k}{\dot T_k|_{21\,\rm cm}}
= \frac{2(1+f_\rmHe)T_k}{\nHI\langle\sigma_{10}v\rangle(T_k)\;
T_\star^2(T_k^{-1}-T_s^{-1})}.
\eeq
Direct computation with the standard evolution of $T_s$ shows that this is 
always much longer than the Hubble time, e.g. $t_h=50\,$Gyr versus 
$H^{-1}=100\,$Myr at $1+z=40$, and $t_h=170\,$Gyr versus $H^{-1}=200\,$Myr 
at $1+z=25$.  A more refined analysis is possible using kinetic theory but 
since the heating timescale is so much longer than any other timescale in 
the problem we need not do this calculation.

\section{Line emissivity and profile}
\label{sec:results}

\subsection{Determination of emissivity}

Having determined the full distribution function for the various levels of 
hydrogen, we now turn to the problem of determining the 21 centimetre 
emissivity.  This emissivity is most conveniently expressed using the time 
derivative of the photon phase space density $\cal N$.  The phase space 
density is related to the familiar specific intensity via
${\cal N} = c^2I_\nu/(2h\nu^3)$ and is more convenient for 
cosmological calculations because, unlike $I_\nu$, it is conserved along a 
trajectory.  Its time derivative is
\beq
\dot{\cal N} = \frac{c^3\rho}{2h\nu^3}(j_\nu-\kappa_\nu I_\nu).
\eeq
In this equation $j_\nu$ comes from spontaneous emission, 
whereas $\kappa_\nu$ comes from a combination of absorption and stimulated 
emission (the latter giving a negative contribution).  The spontaneous 
emission term is given by the usual expression
\beq
j_\nu\rho = \frac{h\nu A}{4\pi}\left[\frac{\rmd\nHI(F=1)}{\rmd 
v_\parallel\,\rmd{\cal V}}\right] \left(\frac{\rmd 
v_\parallel}{\rmd\nu}\right)
= \frac{hcA}{4\pi}\int f_1(\bv)\,\rmd^2\bv_\perp,
\eeq
where the component of velocity along the line of sight is 
$v_\parallel=c(1-\nu/\nu_{10})$, ${\cal V}$ is the volume, we have 
integrated over the irrelevant 
transverse velocity and assumed $|1-\nu/\nu_{10}|\ll 1$.  Since we have 
calculated the distribution function in the bulk rest frame of the gas 
$v_\parallel$ is measured in this frame.  The stimulated 
emission contribution is obtained by 
multiplying this by the photon phase space density ${\cal N}$
and the absorption term is then obtained by replacing 
$f_1\rightarrow f_0$ and multiplying by $3$ to account for the lower
statistical weight of the $F=0$ level:
\beq
-\kappa_\nu\rho I_\nu = \frac{hcA}{4\pi}{\cal N}\int 
[f_1(\bv)-3f_0(\bv)]\,\rmd^2\bv_\perp.
\eeq
We thus find
\beq
\dot{\cal N} = \frac{c^4A}{8\pi\nu^3}\left\{
\int f_1(\bv)\,\rmd^2\bv_\perp
+ {\cal N}\int
[f_1(\bv)-3f_0(\bv)]\,\rmd^2\bv_\perp\right\}.
\eeq
Splitting $f_F(\bv)$ into the thermal piece $f_F^{(\rth)}(\bv)$ and 
non-thermal piece $\xi_F(\bv)$ yields
\beq
\dot{\cal N} = \frac{c^4A}{8\pi\nu^3}\left\{
[y_1(T_k)+{\cal N}y_1(T_k)-3{\cal N}y_0(T_k)]
\nHI\int \Phi_{T_k}(\bv)\,\rmd^2\bv_\perp
+ \int \xi_1(\bv)\,\rmd^2\bv_\perp
+ {\cal N}\int
[\xi_1(\bv)-3\xi_0(\bv)]\,\rmd^2\bv_\perp\right\} \, .
\label{eq:nt}
\eeq
In the limit where $T_\star\ll T_k,T_\gamma$ we have
\beq
y_1(T_k)+{\cal N}y_1(T_k)-3{\cal N}y_0(T_k)
= \frac{
\rme^{T_\star/T_\gamma}(3\rme^{-T_\star/T_k})
- 3
}{(1+3\rme^{-T_\star/T_k})(\rme^{T_\star/T_\gamma}-1)}
\approx \frac34\left(1-\frac{T_\gamma}{T_k}\right).
\eeq
Substituting this into Eq.~(\ref{eq:nt}) and re-expressing it in terms of 
basis modes we find
\beq
\dot{\cal N} = \frac{c^4A}{8\pi\nu^3}\left[
\frac34\left(1-\frac{T_\gamma}{T_k}\right)\nHI\psi_0(v_\parallel)
+ \sum_n \xi_{1n}\psi_n(v_\parallel)
+ {\cal N}\sum_n (\xi_{1n}-3\xi_{0n})\psi_n(v_\parallel)\right],
\eeq
where $\psi_n(v_\parallel)=\int\phi_n(\bv)\,\rmd^2\bv_\perp$ is given by 
the formulas in Appendix~\ref{app:basis}.
Since $\xi'_{\Sigma,n}=0$, we have $\xi_{1n}=\xi'_{\Delta,n}/4$ and since 
${\cal N}\approx T_\gamma/T_\star\gg 1$ the third term dominates 
over the second.  Thus we arrive at
\beq
\dot{\cal N}\approx \frac{c^4A}{8\pi\nu_{10}^3}\left[
\frac34\left(1-\frac{T_\gamma}{T_k}\right)\nHI\psi_0(v_\parallel)
+ \frac{T_\gamma}{T_\star}\sum_n
\xi'_{\Delta,n}\psi_n(v_\parallel)\right].
\label{eq:dndt}
\eeq

The observed brightness temperature $T_b$ due to the \HI\ 21-cm 
signal is obtained by the usual equation, $T_b = h\nu_{\rm 
obs}\Delta{\cal N}/k_B$, where $\nu_{\rm obs}=\nu_{10}/(1+z)$ is the 
observed 
frequency today and $\Delta{\cal N}$ is the change in the phase space 
density.  This gives
\beq
T_b = \frac{h\nu_{10}}{k_B(1+z)}
\Delta{\cal N} = \frac{T_\star}{1+z}
\int \dot{\cal N}\,\rmd t
= \frac{T_\star}
{c(1+z)}\int \dot{\cal N}\,\frac{\rmd r_\parallel}{1+z}
= \frac{T_\star}{c(1+z)^2}\int \dot{\cal N}\,
\rmd r_\parallel,
\label{eq:delta-tb}
\eeq
where $r_\parallel$ is the comoving distance in the radial direction.  In 
this equation $\dot{\cal N}$ should be evaluated using 
Eq.~(\ref{eq:dndt}).  Note that Eq.~(\ref{eq:dndt}) depends on 
$v_\parallel$, which we have defined to be the radial velocity of the H 
atoms that contribute to the signal, relative to the bulk flow of the gas.  
This velocity is
\beq
v_\parallel \approx -\frac{H(z)}{1+z}
[r_\parallel - R_\parallel(z)]-\vpec_\parallel,
\label{eq:vp}
\eeq
where $R_\parallel(z)$ is the radial comoving distance to redshift $z$ in 
an unperturbed universe, $H(z)/(1+z)$ is the cosmological velocity 
gradient, and the peculiar velocity $\vpec_\parallel$ is needed since the 
atom's velocity is measured relative to the gas and not to the Eulerian 
coordinate system.  The negative sign is required since $v_\parallel$ is the 
velocity of the atom relative to the gas (not the other way around) 
although in practice it does not matter since $\psi_n(v_\parallel)$ is even.

\subsection{Large-scale perturbations}

For large-scale perturbations where the velocity gradient and density do 
not vary significantly across the line profile, Eqs.~(\ref{eq:delta-tb}) 
and (\ref{eq:vp}) can be combined to give
\beq
T_b =  \left[
\frac{H(z)}{1+z} + \frac{\rmd \vpec_\parallel}{\rmd r_\parallel}
 \right]^{-1}
\frac{T_\star}{c(1+z)^2}\int \dot{\cal N}\,
\rmd v_\parallel.
\eeq
If we further use Eq.~(\ref{eq:dndt}) for the line profile, we find
\beq
T_b =
\left[
\frac{H(z)}{1+z} + \frac{\rmd \vpec_\parallel}{\rmd r_\parallel}
 \right]^{-1}
\frac{3c^3AT_\star}{32\pi\nu_{10}^3(1+z)^2}\nHI
\left(1-\frac{T_\gamma}{\Tseff}\right),
\label{eq:dtb-new}
\eeq
where the effective spin temperature $\Tseff$ is given by
\beq
\frac1{\Tseff} = 
\frac1{T_k}\int_{-\infty}^\infty \psi_0(v_\parallel)\rmd v_\parallel
- \frac4{3T_\star}\sum_n
\frac{\xi'_{\Delta,n}}{\nHI}\int_{-\infty}^\infty
\psi_n(v_\parallel)\rmd v_\parallel
=
\frac1{T_k}- \frac4{3T_\star}\sum_n
\frac{\sqrt{(2n+1)!}}{2^nn!}
\frac{\xi'_{\Delta,n}}{\nHI}.
\eeq
This result should be compared with Eq.~(\ref{eq:dtb-old}).  Note that 
for the standard assumption of individual Maxwellian distributions for 
$F=0$ and $F=1$ (Eq.~\ref{eq:thermal-xi}), we have 
$\xi'_{\Delta,n}\Tsapprox=-3T_\star(T_s^{-1}-T_k^{-1})\nHI\delta_{n,0}/4$, and 
hence $\Tseff\Tsapprox=T_s$.

For linear-regime perturbations, one may write Eq.~(\ref{eq:dtb-new}) as a 
function $T_b(z, \delta_b, \rmd \vpec_\parallel/\rmd r_\parallel)$.  (We 
consider $T_k$, $\Tseff$, and $\nHI$ to be functions of $\delta_b$.)  
We expand this as
\beq
T_b = \overline{T_b} \left( 1 - \frac{1+z}{H(z)}
\frac{\rmd \vpec_\parallel}{\rmd r_\parallel}\right)
+ \frac{\partial T_b}{\partial\delta_b}\delta_b,
\eeq
where $\bar{T_b}$ is the brightness temperature at mean density with no 
peculiar velocity.  Taking the power spectrum of Eq.~(\ref{eq:dtb-new}),
one may obtain \citep{2005ApJ...624L..65B}
\beq
P_{T_b}(k) = P_{\mu^0}(k) + \mu^2P_{\mu^2}(k) + \mu^4P_{\mu^4}(k)
\eeq
where $\mu=k_\parallel/k$ is the cosine of the angle between the line of 
sight and the wave vector,
\beq
P_{\mu^0}(k) = \left(\frac{\partial T_b}{\partial\delta_b}\right)^2
P_{\delta_b}(k),
\quad
P_{\mu^2}(k) = k\overline{T_b}
\frac{\partial T_b}{\partial\delta_b}
P_{\delta_b,v_b}(k),
\quad{\rm and}\quad
P_{\mu^4}(k) = k^2\left(\overline{T_b}\right)^2 P_{v_b}(k).
\label{eq:pmu}
\eeq
Here $P_{\delta_b}(k)$ is the power spectrum of the baryon density 
fluctuations, $P_{v_b}(k)$ is the power spectrum of the baryon velocity 
perturbations, and $P_{\delta_b,v_b}(k)$ is their 
cross-spectrum.\footnote{\citet{2004MNRAS.352..142B} and 
\citet{2005ApJ...624L..65B} used the additional relation 
$v_b=\delta_b/k$, which is valid for linear evolution on large scales in 
the matter-dominated era.}

\begin{figure}
\includegraphics[width=6in]{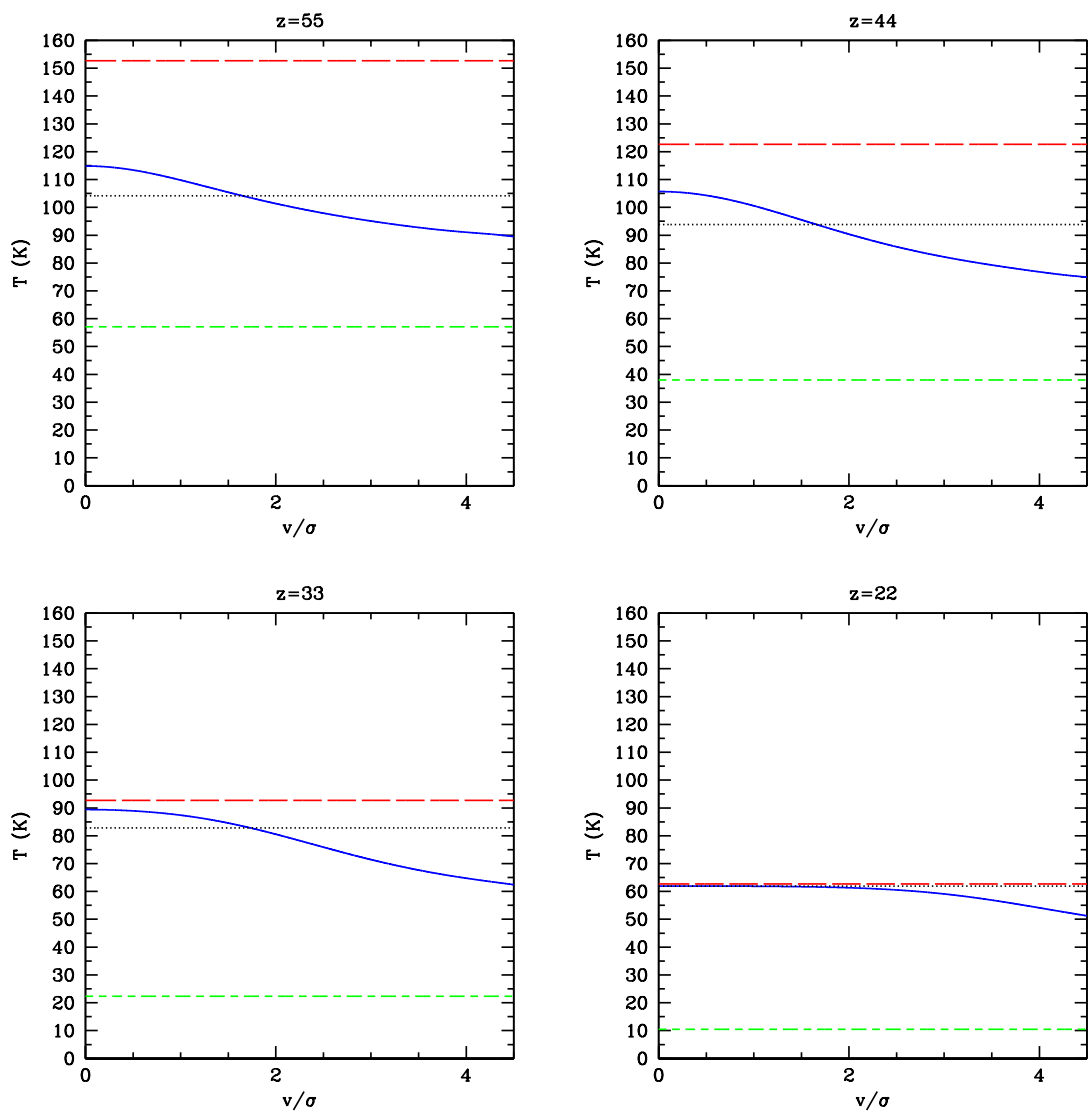}
\caption{\label{fig:ts}The temperature shown for mean 
density intergalactic gas at several redshifts.  The solid (blue) curve 
is the spin temperature as a function of atomic velocity $v$; the 
long-dashed (red) line is the CMB temperature; the short-dashed (green) 
line is the kinetic temperature; and the dotted (black) line is the spin 
temperature computed by the standard formula (Eq.~23 of 
\citealt{1997ApJ...475..429M} with $y_\alpha=0$).  Note that the 
spin temperature is closer to the kinetic temperature for the 
fastest-moving atoms.}
\end{figure}

\begin{figure}
\includegraphics[width=5in]{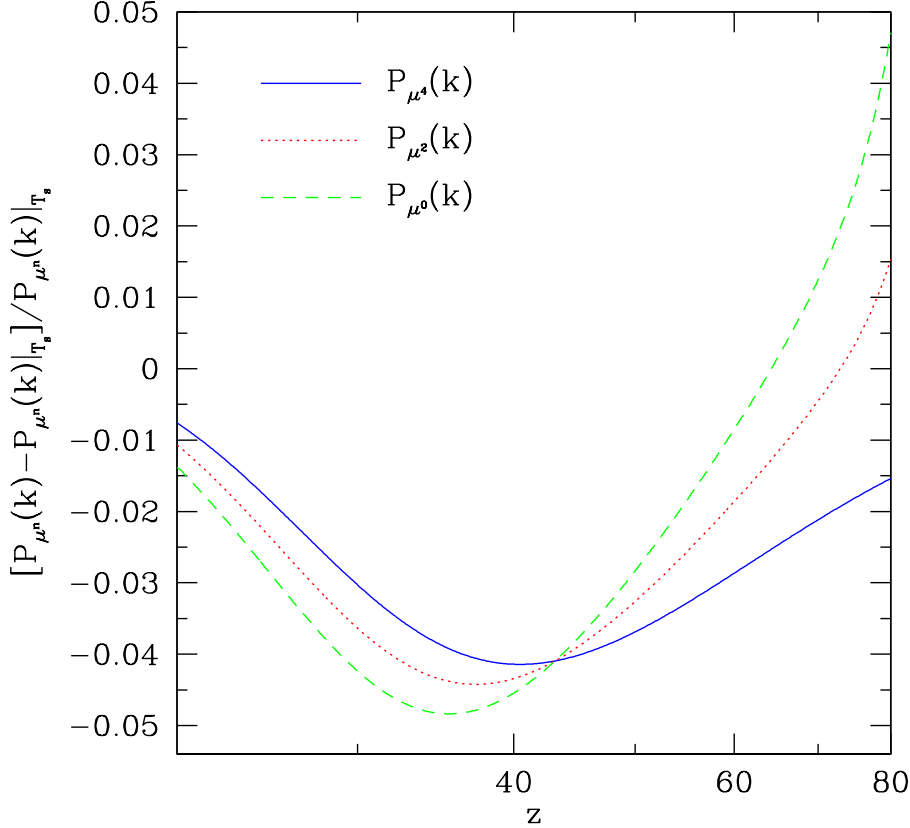}
\caption{\label{fig:relpower}The fractional change in the $\mu^0$, 
$\mu^2$, and 
$\mu^4$ power spectra resulting from the use of the full distribution 
function instead of the standard spin temperature formula (Eq.~23 of
\citealt{1997ApJ...475..429M} with $y_\alpha=0$) versus redshift.}
\end{figure}

\subsection{Results for high redshift intergalactic gas}

We have computed the 21-cm emissivity and line profile for the 
high-redshift intergalactic gas.  The CMB temperature was determined using 
the usual redshift scaling, $T_\gamma=(1+z)T_0$, where $T_0=2.728\,$K is 
the present day temperature \citep{1996ApJ...473..576F}.  The gas kinetic 
temperature was determined using {\sc Recfast} 
\citep{1999ApJ...523L...1S}; we consider only the era before X-ray and 
shock heating become important in controlling the temperature of the gas.  
We do not know precisely when this occurred, but simulations show that 
even in the absence of astrophysical radiation sources, shocked minihaloes 
dominate the {\em mean} 21-cm signal at $z<20$ 
\citep{2005astro.ph.12516S}, and probably dominate the power spectrum even 
sooner.  We therefore cut off our plots at $z=20$ but one should be aware 
that the approximation of adiabatically expanding unshocked gas may break 
down before then.

In Fig.~\ref{fig:ts}, we show the spin temperature of the gas at mean 
density, as a function of the atom's velocity.  For the purposes of this 
figure, we defined a velocity-dependent spin temperature,
\beq
T_s(v) \equiv \frac{T_\star}{\ln [3f_0(v)/f_1(v)]},
\eeq
which is the generalization of the usual definition 
$T_s=T_\star/\ln(3n_0/n_1)$.  We can see that the faster-moving atoms have 
spin temperatures that are closer to the kinetic temperature, which is 
expected since they undergo more frequent collisions.  Conversely, the 
(more typical) slower-moving atoms nearer to the center of the velocity 
distribution have spin temperatures that are closer to the CMB 
temperature.  We thus expect the net effect will be a \emph{reduced} 
magnitude of the 21-cm emissivity.

In Fig.~\ref{fig:relpower}, we show how the power spectrum of 21-cm 
fluctuations on large scales is changed by using kinetic theory in place 
of the usual assumption of a single spin temperature.  These results were 
determined by examining how $\overline{T_b}$ and $\partial
T_b/\partial\delta$ change with the new calculation and then computing 
(c.f. Eq.~\ref{eq:pmu})
\beq
\frac{P_{\mu^0}(k)}{P_{\mu^0}(k)\Tsapprox}
= \left(\frac{[\partial T_b/\partial\delta]}
{[\partial T_b/\partial\delta]\Tsapprox}\right)^2,
\quad
\frac{P_{\mu^2}(k)}{P_{\mu^2}(k)\Tsapprox}
= \frac{[\partial T_b/\partial\delta]}
{[\partial T_b/\partial\delta]\Tsapprox}
\frac{\overline{T_b}}{\overline{T_b}\Tsapprox},
\quad{\rm and}\quad
\frac{P_{\mu^4}(k)}{P_{\mu^4}(k)\Tsapprox}
=\left(
\frac{\overline{T_b}}{\overline{T_b}\Tsapprox}\right)^2.
\eeq
The derivatives with respect to the overdensity $\delta$ were calculated 
assuming the kinetic temperature varies adiabatically, $T_k\propto 
(1+\delta)^{2/3}$.  As one can see, at the low redshifts $z<60$ all 
contributions to the 21-cm signal ($\mu^0$, $\mu^2$, and $\mu^4$) are 
suppressed.  At higher redshift, the \emph{fractional} corrections to 
$P_{\mu^0}(k)$ and $P_{\mu^2}(k)$ become large because $\partial 
T_b/\partial\delta$ crosses through zero at $z\approx 90$.  (The 
zero-crossing occurs because $\partial T_b/\partial\delta$ contains two 
effects: one is the increase in collision rate and overall number of 
hydrogen atoms in overdense regions, which tends to drive $T_b$ to more 
negative values, while on the other hand the increase in $T_k$ means that 
if collisions are efficient, $T_s$ and $T_b$ should go up.  The former 
effect dominates at $z<90$, while the latter effect dominates at $z>90$; 
see e.g. \citealt{2004MNRAS.352..142B}).

\begin{figure}
\includegraphics[width=6in]{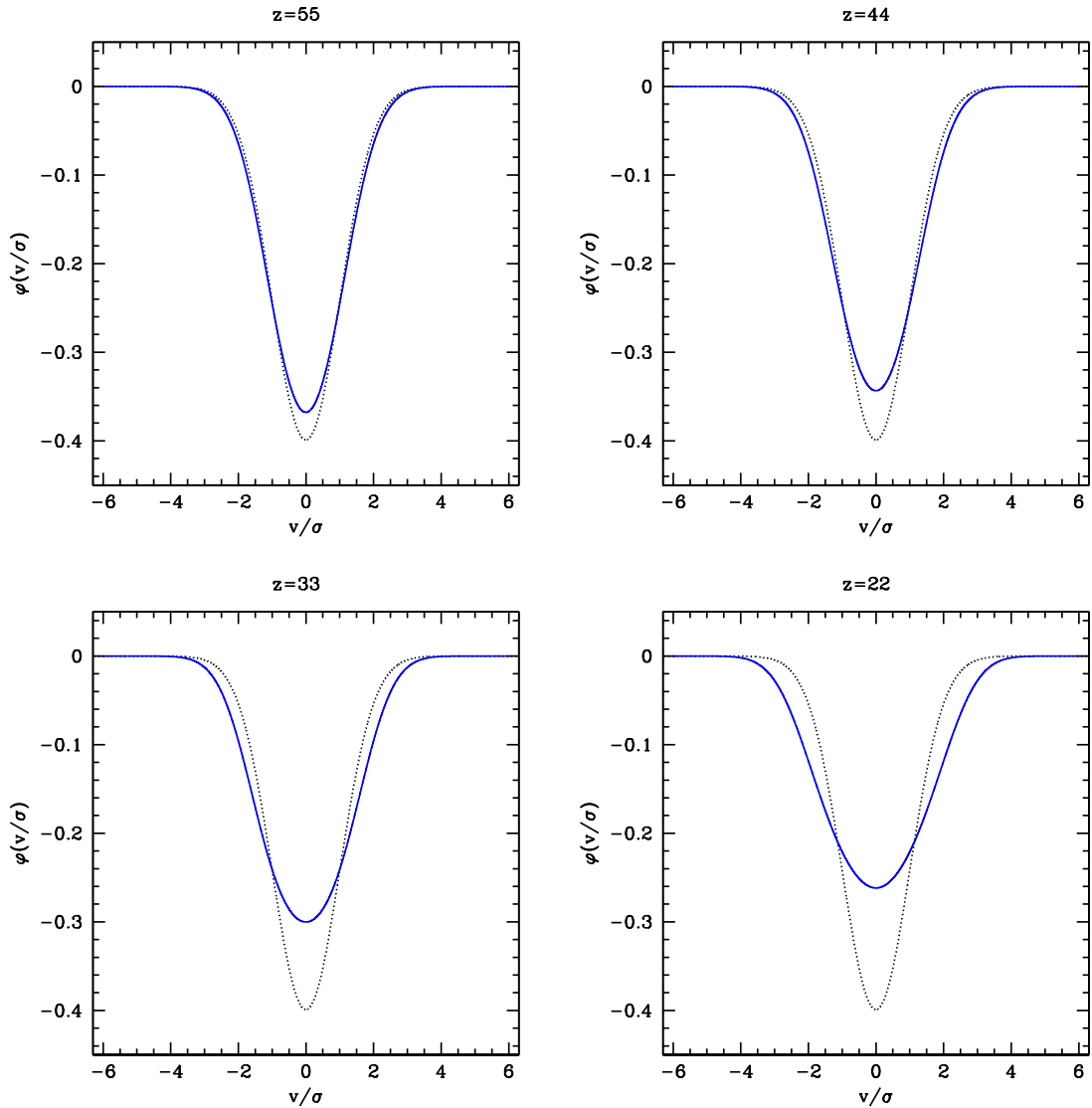}
\caption{\label{fig:lp}The 21-cm line profile shown for mean density 
intergalactic gas at several redshifts.  The solid
(blue) curve is the actual line profile, whereas the dotted (black) curve
shows a Maxwellian profile.  Note that the profile is significantly wider 
than Maxwellian, especially at the lower redshifts.}
\end{figure}

\begin{figure}
\includegraphics[width=6in]{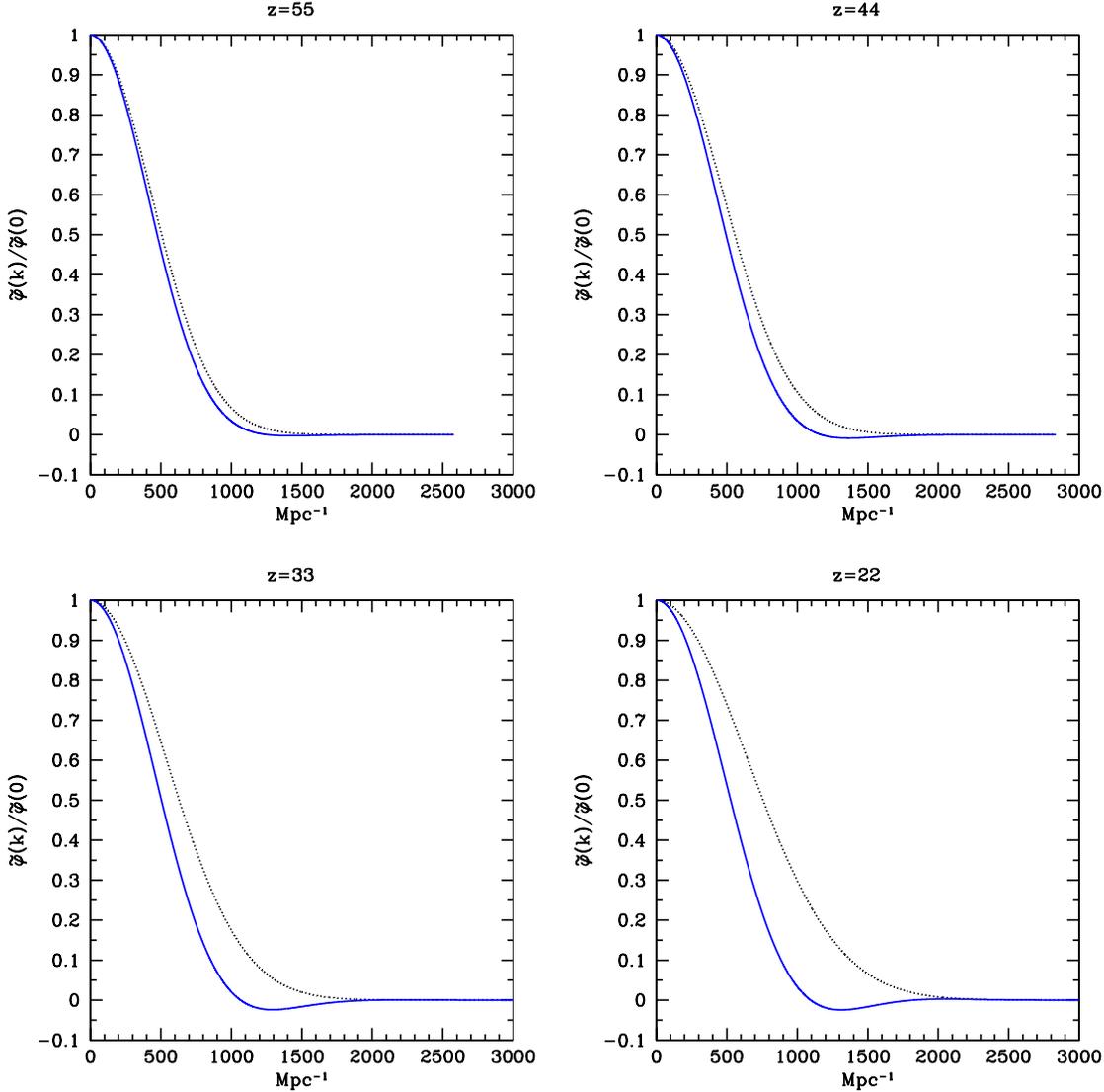}
\caption{\label{fig:tilde}The Fourier transform of the line profile shown 
for mean density intergalactic gas at several redshifts.  The horizontal 
axis has been converted from the conventional km$\,$s$^{-1}$ to the 
redshift space wavenumber, $k_\parallel = (1+z)^{-1}H(z)M_\parallel$.}
\end{figure}

\begin{figure}
\includegraphics[width=5in]{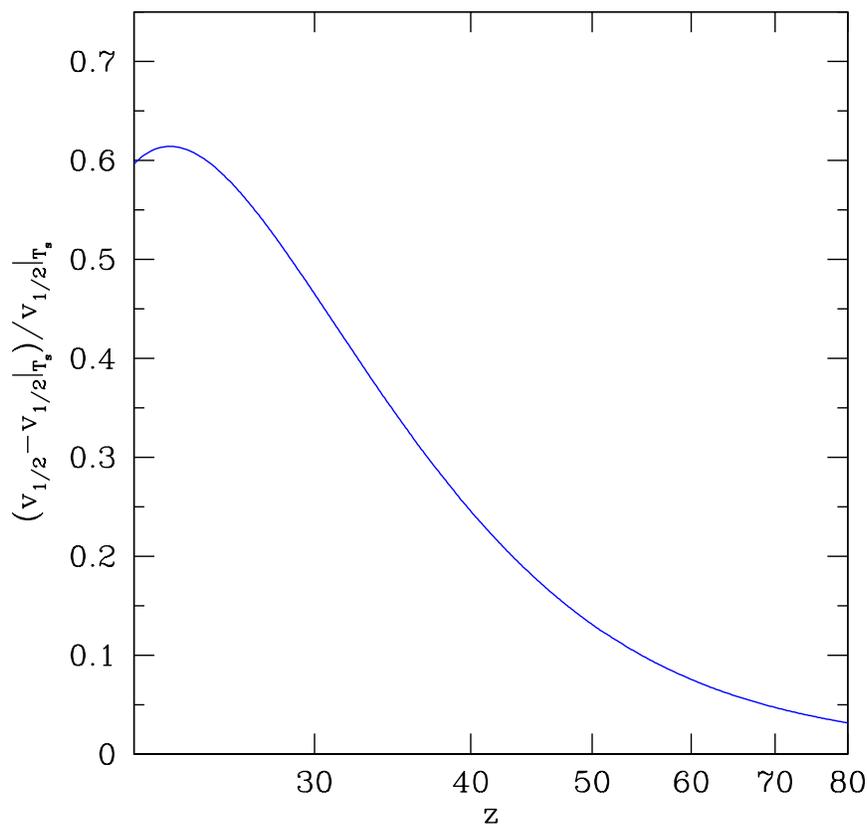}
\caption{\label{fig:v_halfmax}The fractional change in $v_{1/2}$, the 
velocity at half-maximum of the line profile, as a function of redshift.  
Note that in the velocity-independent calculation (for a Gaussian line 
profile) $v_{1/2}\Tsapprox=\sqrt{2 \ln 2}\sigma \simeq 1.18 \sigma$.}
\end{figure}

\begin{figure}
\includegraphics[width=6in]{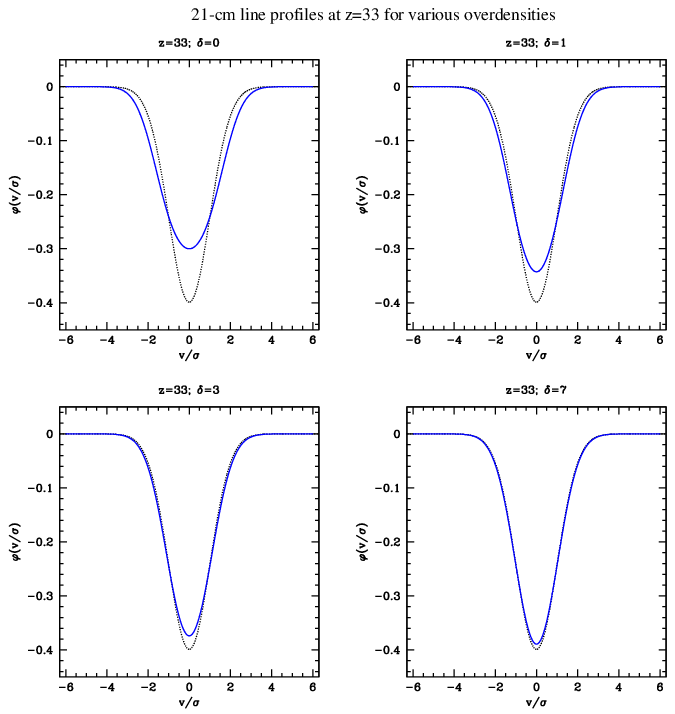}
\caption{\label{fig:33}The 21-cm line profile shown for intergalactic gas 
at $z=33$.  We show values ranging from mean density ($\delta=0$, left 
panel) through 8 times mean density ($\delta=7$, right panel), assuming a 
kinetic temperature from adiabatic evolution, 
$T_k\propto(1+\delta)^{2/3}$.  The solid (blue) curve is the actual line 
profile, whereas the dotted (black) curve shows a Maxwellian profile.  
Note that the line profile is more nearly Gaussian at high densities and 
temperatures where collisions are more effective.}
\end{figure}

The 21-cm line profile, $\varphi(v)$, is needed if one wishes to consider 
the smallest-scale fluctuations in the 21-cm radiation.  The line profile 
is a function of the radial velocity and is proportional to the rate of 
photon emission, i.e. $\varphi(v)\propto \dot{\cal N}$ (see 
Eq.~\ref{eq:dndt}).  The line profiles for gas at mean density are shown 
in Fig.~\ref{fig:lp}; here the line profiles have been normalized
such that $|\int_{-\infty}^{\infty} \varphi(x)\,\rmd x| = 1$ (where
$x=v/\sigma$).  Note that they are wider than Maxwellian, because the 
fast-moving atoms have a lower spin temperature (i.e. farther from 
$T_\gamma$) than the slow-moving atoms.  In Fig.~\ref{fig:tilde} we show 
the Fourier transform of the line profile, which determines the Doppler 
cutoff of the 21-cm power spectrum in the radial direction.  In the 
Maxwellian case, the Fourier transform is simply a Gaussian,
\beq
\frac{\tilde\varphi(k)}{\tilde\varphi(0)} = \rme^{-k_\parallel^2/2k_T^2},
\eeq
where $k_T=(1+z)^{-1}H(z)\sigma^{-1}$. Unsurprisingly, the 
wider-than-Maxwellian line profile in velocity space manifests itself as a 
cutoff at smaller radial wavenumber $k_\parallel$.  This leads to an 
additional suppression of modes in the 21-cm signal with $|k_\parallel|$ 
greater than a few hundred Mpc$^{-1}$, which unfortunately makes them even 
harder to observe.  It also means that proposals to measure the IGM 
temperature using the redshift-space anisotropy caused by the thermal 
motions of the atoms \citep{2005MNRAS.362.1047N} will have to consider in 
detail the non-Maxwellian nature of the line profile.  The widening of the 
line profile can also be illustrated by plotting the increase in FWHM 
relative to the Maxwellian case; this is shown in 
Fig.~\ref{fig:v_halfmax}.  The line profile for adiabatically compressed 
gas is shown in Fig.~\ref{fig:33}.  Note that the high-density gas 
(e.g. in the bottom-right panel with $1+\delta=8$)
has a nearly Maxwellian line profile, as expected when collisions dominate 
over radiative transitions.

\section{Discussion}
\label{sec:dis}

The redshifted 21-cm radiation from the dark ages is a promising future 
cosmological probe, and in recent years there has been substantial 
progress on understanding the theory of this radiation.  However in the 
analyses to date, it has been assumed that the \HI\ atoms can be described 
by a single spin temperature $T_s$ and a kinetic temperature $T_k$.  We 
have removed this assumption by describing the gas with a full joint 
spin-velocity distribution function $f_F(v)$.  We find that although the 
overall (spin-summed) velocity distribution is Maxwellian, the individual 
velocity distributions of \HI\ in the excited and de-excited hyperfine 
levels are not.  This leads to two effects: a widening of the 21-cm line,
i.e. width greater than $\sqrt{k_BT_k/\mH}$; and a suppression of the 
overall 21-cm signal.

The suppression of the signal on large scales amounts to an effect of up 
to $\sim\!5$ per cent in the power spectrum.  While this is small, it will 
be essential to consider it if precision cosmological constraints are ever 
to be obtained from the pre-reionization 21-cm signal 
\citep{2004PhRvL..92u1301L, 2005MNRAS.362.1047N}.  On small scales 
($\sim\!10^{-2}\,$Mpc) the 21-cm line may be broadened up to $\sim\!1.6$ 
times its Maxwellian width.  However, observing these very-small scale 
features may prove to be even more difficult than the large-scale features 
since the amount of power per mode is less on small scales, where 
$P(k)\propto k^{-3}$.  An alternative method of identifying small-scale 
structures is to look at 21-cm absorption features in the spectra of very 
high-redshift radio sources \citep{2002ApJ...577...22C, 
2002ApJ...579....1F}, but it is very unlikely that such sources will be 
found behind a collisionally spin-coupled IGM except in models with 
unusually early structure formation.  Another way that the line profile 
could be important is in regions such as minihaloes that have significant 
optical depth \citep{2005astro.ph.12516S}, since line self-absorption is 
suppressed when the line is broadened.  In this case the contribution of 
the minihaloes to the large-scale power spectrum could in principle be 
detectable even if individual minihaloes are not identified, although even 
a small amount of \lya\ emission could decouple the IGM spin temperature 
from the CMB and cause the IGM to overwhelm the minihalo signal 
\citep{2006astro.ph..4080F}.  Note that in the case of a radio source or 
minihalo, the absorption/emission line profile would be modified since one 
is no longer seeing the gas against a background at brightness temperature 
$T_\gamma$.

At some point in the evolution of the Universe, UV sources turn on and the 
IGM temperature is affected by \lya\ scattering as well as by collisional 
and 21-cm transitions (see \citealt{2006astro.ph..4040F} for some recent 
models).  A detailed analysis of the spin-velocity distribution in this 
case is beyond the scope of this paper, although we suspect that the 
single spin temperature approximation is very good for this case.  This is 
because, at temperatures greater than a few Kelvin and typical IGM \lya\ 
optical depths (of order $10^6$), the colour temperature $T_c$ of the 
\lya\ photons relaxes to the gas kinetic temperature, and the width of 
features in the \lya\ spectrum is much wider than the Doppler shift 
frequency $\sim \nu_\lya\sigma/c$ induced by the motions of the atoms 
\citep{2004ApJ...602....1C}; thus all atoms see a similar \lya\ radiation 
field at similar colour temperature.  This circumstance would of course 
lead to a velocity-independent spin temperature if \lya\ and 21-cm 
dominate over collisions.  At temperatures of several Kelvin, obtainable 
in low-density regions if reionization starts late, the colour temperature 
can exhibit more complicated behaviour if $T_s\neq T_k$ 
\citep{2005astro.ph..7102H}.  This could potentially produce a 
non-Maxwellian line profile and affect the total emissivity; in order to 
solve this problem one would need to simultaneously solve the radiative 
transfer and Boltzmann equations for the UV photon spectrum and \HI\ 
velocity distribution.  However the single spin temperature approximation 
would apply to high accuracy if X-ray heating drives $T_k$ to tens of K or 
more, or if the \lya\ spin-flip scattering rate ever becomes much larger 
than the 21-cm transition rate (which would thermalize all spins at 
$T_k$). Collisions with electrons may also be important if there is 
partial ionization due to X-rays \citep{2006ApJ...637L...1K}; this 
hyperfine transition mechanism is expected to be independent of the atom's 
velocity because this is negligible compared to the velocity of the 
electron.

We conclude that the traditional assumption of a single spin temperature 
describing the high-redshift gas, while a good first approximation, is
incorrect at the several per cent level for the mean signal and at the 
several tens of per cent level for the line profile.  Any calculation 
aiming for higher precision must thus include the full kinetic theory 
analysis and the joint spin-velocity distribution of the neutral hydrogen 
atoms derived here.

\section*{Acknowledgments}
We thank S. Furlanetto for insightful comments on a draft of this paper. 
CH is supported in part by NSF PHY-0503584 and by a grant-in-aid from the 
W. M. Keck Foundation. KS is supported by NASA through Hubble Fellowship 
grant HST-HF-01191.01-A awarded by the Space Telescope Science Institute, 
which is operated by the Association of Universities for Research in 
Astronomy, Inc., for NASA, under contract NAS 5-26555.

\appendix

\section{Properties of the basis functions}
\label{app:basis}

Here we summarize some useful properties of the basis functions defined in 
Eq.~(\ref{eq:phi-n}).  These are needed in order to compute the two 
quantities of observational interest, namely the velocity-marginalized 
\HI\ level populations $\{n_0,n_1\}$ and the 21-cm line profile (which in 
general can be non-Gaussian).

Both the level population integrals and the line profile are related to 
the 3-dimensional Fourier transform of $\phi_n$, namely 
$\tilde\phi_n(\bM)=\int \phi_n(\bv) \rme^{-\rmi \bM\cdot\bv}\,\rmd^3\bv$.  
This quantity can be evaluated by switching to spherical coordinates for 
$\bv$:
\beq
\tilde\phi_n(\bM) = \int \phi_n(\bv) \rme^{-\rmi Mv\cos\theta}
 \,v^2\,\sin\theta\,\rmd v\,\rmd\theta\,\rmd\varphi
= 4\pi \int_0^\infty \phi_n(\bv) \frac{\sin (Mv)}{Mv} v^2\,\rmd v.
\eeq
Recalling that $\sigma=\sqrt{k_BT_k/\mH}$, we then have, using 
Eq.~(\ref{eq:phi-n}),
\beq
\tilde\phi_n(\bM) = \frac{2^{-n-1/2}\pi^{-1/2}[(2n+1)!]^{-1/2}}{\sigma^2M}
\int_0^\infty H_{2n+1}\left(\frac v\sigma\right)\rme^{-v^2/2\sigma^2}
\sin(Mv)\,\rmd v.
\eeq
We note that $H_{2n+1}$ is odd so the integrand here is even; thus we may 
replace $\int_0^\infty\rightarrow\frac12\int_{-\infty}^\infty$.  Also
$\sin(Mv)=\Im\rme^{\rmi Mv}$ where $\Im$ denotes imaginary part.  If we 
further substitute $x=v/\sigma$, we get
\beqa
\tilde\phi_n(\bM) &=& \frac{2^{-n-3/2}\pi^{-1/2}[(2n+1)!]^{-1/2}}{\sigma 
M}\Im\int_{-\infty}^\infty H_{2n+1}(x)
\rme^{-x^2/2} \rme^{\rmi M\sigma x}\,\rmd x
\nonumber \\
&=& \frac{2^{-n-1}[(2n+1)!]^{-1/2}}{\sigma M}
(-1)^n \rme^{-\sigma^2M^2/2} H_{2n+1}(\sigma M),
\label{eq:tpn}
\eeqa
where in the second equality we have used the Fourier transform of a 
product of a Hermite polynomial and a Gaussian (e.g. Eq.~7.376.1 of 
\citealt{1994tisp.book.....G}).  Using the specific equation for 
$H_{2n+1}$, we can see that the integrated line profile is
\beq
\tilde\phi_n(0) = (-1)^n2^{-n-1}[(2n+1)!]^{-1/2}
\lim_{M\rightarrow 0} \frac{H_{2n+1}(\sigma M)}{\sigma M}
= (-1)^n2^{-n-1}[(2n+1)!]^{-1/2}H'_{2n+1}(0)
\eeq
according to l'H\^{o}pital's rule.  The derivative $H'_{2n+1}(0)$ can be 
determined from the series expansion of $H_{2n+1}$ and gives
\beq
H'_{2n+1}(0) = (-1)^n\frac{(2n+2)!}{(n+1)!}
\quad\rightarrow\quad
\tilde\phi_n(0) = \frac{\sqrt{(2n+1)!}}{2^n\,n!}.
\label{eq:AR1}
\eeq

In addition to the integrals of the basis modes, provided by 
Eq.~(\ref{eq:AR1}), we also need the contribution of each basis mode to 
the line profile, defined by
\beq
\psi_n(v_\parallel) = \int \phi_n(\bv)\,\rmd^2\bv_\perp,
\eeq
where $v_\parallel$ is the component of $\bv$ along the line of sight and 
$\bv_\perp$ is the component in the plane of the sky.  The Fourier 
transform of the line profile is trivially determined,
\beqa
\tilde\psi_n(M_\parallel) &\equiv&
\int_{-\infty}^\infty \psi_n(v_\parallel)
\rme^{\rmi M_\parallel v_\parallel} \,\rmd v_\parallel
= \int \phi_n(\bv) \rme^{\rmi M_\parallel v_\parallel}\,\rmd 
v_\parallel\,\rmd^2\bv_\perp = \tilde\phi_n(M_\parallel)
\nonumber \\ &=&
\frac{2^{-n-1}[(2n+1)!]^{-1/2}}{\sigma M_\parallel}
(-1)^n \rme^{-\sigma^2M_\parallel^2/2} H_{2n+1}(\sigma M_\parallel).
\label{eq:AR2}
\eeqa
It is in fact this Fourier transform $\tilde\psi_n(M_\parallel)$ that we 
will need in order to obtain linear theory power 
spectra.  However it is conceptually useful to plot the 
actual function $\psi_n(v_\parallel)$, so we calculate this here.  It is
\beq
\psi_n(v_\parallel) = \int_0^\infty \phi_n\left(
\sqrt{v_\parallel^2+v_\perp^2}\right) \, 2\pi v_\perp\,\rmd v_\perp
= 2\pi \int_{|v_\parallel|}^\infty \phi_n(v)\,v\,\rmd v
= 2\pi \int_{v_\parallel}^\infty \phi_n(v)\,v\,\rmd v.
\eeq
Here the second equality involves the change of variables from $v_\perp$ 
to $v=\sqrt{v_\parallel^2+v_\perp^2}$, and the third equality holds
because the integrand is odd and hence the integral from $v_\parallel$ to 
$|v_\parallel|$ contributes nothing.  Plugging in the specific form for 
the basis modes from Eq.~(\ref{eq:phi-n}) gives
\beq
\psi_n(v_\parallel) = \frac1{2\sqrt{2\pi(2n+1)!}\,\sigma^2}
\int_{v_\parallel}^\infty H_{2n+1}\left(\frac v\sigma\right)
\rme^{-v^2/2\sigma^2}\,\rmd v
= \frac1{2\sqrt{2\pi(2n+1)!}\,\sigma}
\int_{v_\parallel/\sigma}^\infty H_{2n+1}(x)
\rme^{-x^2/2}\,\rmd x,
\label{eq:psin-parallel}
\eeq
where we have again substituted $x=v/\sigma$.  The integral may be solved 
by using the Hermite polynomial recurrence relations (Eqs. 8.952.1 and 
8.952.2 of \citealt{1994tisp.book.....G}) in the form
\beqa
\frac\rmd{\rmd x}\left[\frac1{2^{2j-1}j!}H_{2j}(x)\rme^{-x^2/2}\right]
&=& \frac1{2^{2j-1}j!}[H'_{2j}(x)-xH_{2j}(x)]\rme^{-x^2/2}
= 
\frac1{2^{2j-1}j!}\left[2jH_{2j-1}(x)-\frac12H_{2j+1}(x)\right]\rme^{-x^2/2}
\nonumber \\
&=&
\left[ \frac{H_{2j-1}(x)}{2^{2(j-1)}(j-1)!}
- \frac{H_{2j+1}(x)}{2^{2j}j!} \right] \rme^{-x^2/2}
\eeqa
for $j\ge 1$.  In the special case of $j=0$, it is trivial to see that 
the same expression is valid if one formally defines 
$H_{-1}(x)/[(-1)!]=0$.  Summing this equation 
from $j=0$ to $n$, we then get
\beq
\frac\rmd{\rmd x}\left[\sum_{j=0}^n 
\frac1{2^{2j-1}j!}H_{2j}(x)\rme^{-x^2/2}\right]
= -\frac{H_{2n+1}(x)}{2^{2n}n!}\rme^{-x^2/2}.
\eeq
With this equation, we can now solve the integral in 
Eq.~(\ref{eq:psin-parallel}),
\beq
\int_{v_\parallel/\sigma}^\infty H_{2n+1}(x)
\rme^{-x^2/2}\,\rmd x = -2^{2n}n!
\sum_{j=0}^n\frac1{2^{2j-1}j!}
\left.H_{2j}(x)\rme^{-x^2/2}\right|_{x=v_\parallel/\sigma}^\infty
= \sum_{j=0}^n 
2^{2(n-j)+1}\frac{n!}{j!}H_{2j}\left(\frac{v_\parallel}\sigma
\right)
\rme^{-v_\parallel^2/2
\sigma^2},
\eeq
which implies
\beq
\psi_n(v_\parallel) = \frac1{\sqrt{2\pi}\,\sigma}\,
\frac{2^nn!}{\sqrt{(2n+1)!}}\sum_{j=0}^n \frac{H_{2j}(v_\parallel/\sigma)}
{2^{2j}j!}\rme^{-v_\parallel^2/2\sigma^2}.
\label{eq:AR3}
\eeq
This formula is suitable for numerical evaluation and has been used to 
generate our plots of the 21-cm line profile.

\section{Cross Sections}
\label{app:cross}

\subsection*{Derivation of H-H Cross Section}

We determine the H-H scattering cross section according to the elastic 
approximation \citep{1961PRSLA.262..132D}.  In this approximation we 
ignore the hyperfine energy defect and treat the collision as a scattering 
problem with separate potentials $V_{0}(R)$ and $V_{1}(R)$ for the 
electron spin-singlet ($S=0$) and spin-triplet ($S=1$) states.  The 
difference between these potentials results in a change in the electronic 
spin of a particular hydrogen atom, and hence a change in its total 
(nuclear plus electronic) spin angular momentum.  The derivation below is 
based on the formalism of \citet{1966P&SS...14..929S, 
1966P&SS...14..937S}, although we use the $|SIF_tM_{F_t}\rangle$ basis 
rather than diagonalizing in $S$ and $M_S$ as was done by 
\citet{1966P&SS...14..929S}.

There have been many other previous determinations of the cross section 
for scattering of two hydrogen atoms, some going beyond the elastic 
approximation; however the published results do not provide the full spin 
and angular dependence of the cross section, tabulating instead the 
spin-flip cross section \citep{1969ApJ...158..423A, 2003PhRvA..67d2715Z, 
2005ApJ...622.1356Z} or moments of the angular distribution 
\citep{1992PhRvA..46.6956J, 2000PhRvA..61a4701J}.

We are interested in the differential cross section for hydrogen atoms in 
the $F'$ and $F''$ hyperfine levels to scatter and leave a hydrogen atom 
in the $F$ level moving in direction $\hat\bOmega$.  It will be assumed 
that initially the $F''$ atom is moving with velocity $w\be_3$ relative to 
the $F'$ atom, so that $\theta=\arccos\hat\Omega_3$ is the scattering 
angle.  In our problem the hydrogen spins are unpolarized because the 
radiation field is isotropic and because in the elastic approximation the 
atoms cannot be polarized by collisions.  Therefore we consider only the 
cross section summed over final magnetic quantum number $M_F$ and averaged 
over initial $M_{F'}$ and $M_{F''}$.

We denote
the total spin by $F_t$ (the vector sum of $F'$ and $F''$), and use the 
labels $a$ and $b$ for the nuclei.  The incident wave function of the two 
hydrogen atoms is then
\beq
|\Psi_{\rm in}\rangle = {\cal A}_p{\cal A}_e \left(
|1s_a1s_b\rangle |F'F''F_tM_{F_t}\rangle \rme^{ikZ} \right),
\label{eq:psi-in}
\eeq
where $(X,Y,Z)=\bR_b-\bR_a$ is the relative position vector of the two 
nuclei, $k=\mH w/2\hbar$ is the wavevector for reduced mass $\mH/2$, 
$|1s_a1s_b\rangle$ is the orbital wave function of the two electrons (with 
$1s_a$ and $1s_b$ representing the $1s$ orbitals associated with nuclei 
$a$ and $b$), $|F'F''F_tM_{F_t}\rangle$ is the spin state of the nuclei 
$a$ and 
$b$ and electrons $1$ and $2$, and ${\cal A}_{p,e}$ are the 
antisymmetrization operators for the protons and electrons.  We have 
chosen our coordinate system such that the relative velocity is along the 
third coordinate axis.  Also the notation $|F_{a1}F_{b2}F_tM_{F_t}\rangle$ 
refers to the angular momenta ${\bmath F}_{a1} = {\bmath I}_a+{\bmath 
S}_1$ and ${\bmath F}_{b2}={\bmath I}_b+{\bmath S}_2$.

The antisymmetrization operators in Eq.~(\ref{eq:psi-in}) commute with 
the total electronic and nuclear angular momenta $S$ and $I$, but not with 
$F_{a1}$ and $F_{b2}$.  It is thus convenient to re-write 
Eq.~(\ref{eq:psi-in}) in the basis of eigenstates of $S$ and $I$ (which 
we denote $|SI,F_tM_{F_t}\rangle$):
\beqa
|\Psi_{\rm in}\rangle &=& {\cal A}_p{\cal A}_e \left(
|1s_a1s_b\rangle
\sum_{S,I}
\langle SI,F_tM_{F_t}|F'F''F_tM_{F_t}\rangle
 |SI,F_tM_{F_t}\rangle
 \rme^{ikZ} \right)
\nonumber \\ &=&
\frac{1}{2}
\sum_{S,I}
[|1s_a1s_b\rangle + (-1)^S |1s_b1s_a\rangle]
\langle SI,F_tM_{F_t}|F'F''F_tM_{F_t}\rangle
|SI,F_tM_{F_t}\rangle
[\rme^{ikZ} + (-1)^{S+I}\rme^{-ikZ}].
\label{eq:psi-in-si}
\eeqa

The scattered wave function is related to the ingoing wave function by the 
scattering amplitudes $f(\hat\bOmega)$.  These are defined in terms of the 
large-$R$ asymptotic form for the wave function,
\beq
|\Psi\rangle \sim \rme^{ikZ} + \frac{\rme^{ikR}}{R}f(\hat\Omega),
\label{eq:psi}
\eeq
in which an incident state $|\Psi_{\rm 
in}\rangle \sim \rme^{ikZ}$ is mapped into an outgoing scattered state 
$|\Psi_{\rm 
scat}\rangle \sim R^{-1}\rme^{ikR}f(\hat\bOmega)$.    In our case, the 
ingoing wave function has 
both singlet and triplet parts with different potentials and hence 
different scattering amplitudes.  Thus we must superpose singlet and 
triplet solutions in order to obtain the wave function corresponding to 
Eq.~(\ref{eq:psi-in-si}).  Since our particles are identical, we must also 
superpose both the ingoing wave function with $\rme^{ikZ}$ dependence, and 
one with $\rme^{-ikZ}$ dependence which can be obtained from 
Eq.~(\ref{eq:psi}) by a parity transformation.  Thus the wave 
function of Eq.~(\ref{eq:psi-in-si}) yields an outgoing scattered state
\beq
|\Psi_{\rm scat}\rangle =
\frac{\rme^{ikR}}{2R}
\sum_{S,I}
[|1s_a1s_b\rangle + (-1)^S |1s_b1s_a\rangle]
\langle SI,F_tM_{F_t}|F'F''F_tM_{F_t}\rangle
|SI,F_tM_{F_t}\rangle
[f_S(\hat\bOmega) + (-1)^{S+I}f_S(-\hat\bOmega)],
\eeq
where the amplitudes $f_0(\hat\bOmega)$ and $f_1(\hat\bOmega)$ pertain to 
electron spin singlets and triplets respectively.  These scattering 
amplitudes are determined by solving the Schr\"{o}dinger equation by the 
usual partial wave method and we describe the potentials used and other 
pertinent details at the end of this Appendix.

The differential scattering cross section to put 
nucleus $a$ and electron $1$ into an H atom with total angular momentum 
$F$ in the final state is then obtained by projecting $|\Psi_{\rm 
scat}\rangle$ onto all outgoing states with $F_{a1}=F$ and arbitrary 
$F_{b2}$, and then summing the square norms:
\beqa
\frac{\rmd\sigma}{\rmd\hat\bOmega}(F_{a1}=F)
&=& R^2
\sum_{F_{b2}} \left| \langle FF_{b2}F_tM_{F_t}|\Psi_{\rm scat}\rangle
\right|^2
\nonumber \\
&=& \frac{1}{4}\sum_{F_{b2}} \left|
\sum_{S,I}
\langle SI,F_tM_{F_t}|F'F''F_tM_{F_t}\rangle
\langle FF_{b2}F_tM_{F_t}|SI,F_tM_{F_t}\rangle
[f_S(\hat\bOmega) + (-1)^{S+I}f_S(-\hat\bOmega)]
\right|^2.
\eeqa
(In principle we must also sum over the final values of the $F_t$ and 
$M_{F_t}$ quantum numbers; however since we have neglected interactions 
involving the spin these will be conserved and we may simply use their 
initial values.)  To get our final result for the differential cross 
section for producing a hydrogen atom with total spin $F$ moving in the 
direction $\hat\bOmega$, we must average over the initial values of the 
quantum numbers $F_t$ and $M_{F_t}$ in their statistical ratios, and then 
multiply by 4 because the final-state hydrogen atom could contain either 
nucleus ($a$ or $b$) and either electron ($1$ or $2$).  The summation over 
$M_{F_t}$ is trivial since none of the inner products or amplitudes inside 
the sum depend on it:
\beqa
\sigma_{F'F''}\frac{\rmd P_{F|F'F''}}{\rmd\hat\bOmega}
&=& \frac{1}{(2F'+1)(2F''+1)} \sum_{F_t,F_{b2}} (2F_t+1) \Biggl|
\sum_{S,I}
\langle SI,F_tM_{F_t}|F'F''F_tM_{F_t}\rangle
\langle FF_{b2}F_tM_{F_t}|SI,F_tM_{F_t}\rangle
\nonumber \\ && \times
[f_S(\hat\bOmega) + (-1)^{S+I}f_S(-\hat\bOmega)]
\Biggr|^2.
\label{eq:sigfff}
\eeqa

The evaluation of Eq.~(\ref{eq:sigfff}) is a straightforward but tedious 
exercise in angular momentum recoupling theory.  The recoupling 
coefficients $\langle SI,F_tM_{F_t}|F'F''F_tM_{F_t}\rangle$ are given in 
terms of the 9-$j$ symbol by
\beq
\langle SI,F_tM_{F_t}|F'F''F_tM_{F_t}\rangle = 
\sqrt{(2S+1)(2I+1)(2F'+1)(2F''+1)} \left\{\begin{array}{ccc}
1/2 & 1/2 & S \\ 1/2 & 1/2 & I \\ F' & F'' & F_t
\end{array}\right\}
\eeq
\citep{1960amqm.book.....E}.  Using the specific values of the 9-$j$ 
symbols, we find
\beqa
\sigma_{00}\frac{\rmd P_{0|00}}{\rmd\hat\bOmega}
&=& \frac{1}{16}|\fzz+3\foo|^2,
\nonumber \\
\sigma_{00}\frac{\rmd P_{1|00}}{\rmd\hat\bOmega}
&=& \frac{3}{16}|\fzz-\foo|^2,
\nonumber \\
\sigma_{01}\frac{\rmd P_{0|01}}{\rmd\hat\bOmega}
&=& \frac{1}{16}|\fzo+\foz+2\foo|^2,
\nonumber \\
\sigma_{01}\frac{\rmd P_{1|01}}{\rmd\hat\bOmega}
&=& \frac{1}{16}|\fzo+\foz-2\foo|^2 + \frac{1}{8}|\fzo-\foz|^2,
\nonumber \\
\sigma_{10}\frac{\rmd P_{0|10}}{\rmd\hat\bOmega}
&=& \frac{1}{16}|\fzo+\foz-2\foo|^2,
\nonumber \\
\sigma_{10}\frac{\rmd P_{1|10}}{\rmd\hat\bOmega}
&=& \frac{1}{16}|\fzo+\foz+2\foo|^2 + \frac{1}{8}|\fzo-\foz|^2,
\nonumber \\
\sigma_{11}\frac{\rmd P_{0|11}}{\rmd\hat\bOmega}
&=& \frac{1}{48}|\fzz-\foo|^2 + \frac{1}{24}|\fzo-\foz|^2,
\quad{\rm and}
\nonumber \\
\sigma_{11}\frac{\rmd P_{1|11}}{\rmd\hat\bOmega}
&=& \frac{1}{144}|3\fzz+\foo|^2+\frac{1}{24}|\fzo-\foz|^2
  + \frac{1}{12}|\fzo+\foz|^2 + \frac{5}{9}|\foo|^2,
\label{eq:sigma-H}
\eeqa
where $f_{SI}(\hat\bOmega)=f_S(\hat\bOmega)+(-1)^{S+I}f_S(-\hat\bOmega)$.  

\subsection*{Derivation of He-H Cross Section}

The He-H cross section is actually simpler to derive than H-H because the 
He atom has no electronic spin or orbital angular momentum.  Thus there is 
only one relevant electronic state, $X^2\Sigma^+$, and there is no spin 
exchange.  The scattering cross section is independent of spin and given 
by the usual formula, $\rmd\sigma/\rmd\hat\bOmega = |f(\hat\bOmega)|^2$, 
where $f(\hat\bOmega)$ is the scattering amplitude in the $X^2\Sigma^+$ 
potential.

\subsection*{Potentials, Phase Shifts and Scattering Amplitudes}

In order to calculate the cross sections as outlined above we must 
calculate the scattering amplitude $f(\hat\bOmega)$ in the atomic 
potential $V(R)$. To do this we solve the Schr\"{o}dinger equation
\beq
-\frac{\hbar^2}{2\mu}\nabla^2\Psi+(V-E)\Psi = 0 \, ,
\eeq
where $\Psi=\langle \bR | \Psi \rangle$, $\mu$ is the reduced mass of the 
two colliding atoms, and $E=\hbar^2k^2/2\mu$  for an unbound state with 
wavevector $\bk$.  For a spherically symmetric potential the wavefunction 
can be expanded in
a basis of angular momentum eigenfunctions as
\beq
\Psi(R,\hat\bOmega)=\sum_{N=0}^{\infty} \frac{a_N}{R} 
\psi_N(R)P_N(\hat\bk\cdot\hat\bOmega)
\eeq
where $N$ is the orbital angular momentum quantum number,
the $a_N$ are constant coefficients, $P_N$ are Legendre 
polynomials, and the partial wave $\psi_N$ satisfies the radial equation
\beq
\label{eq:rad}
\frac{\rmd^2 \psi_N}{\rmd 
R^2}+\left[k^2-\frac{N(N+1)}{R^2}-\frac{2\mu}{\hbar^2}V(R)\right]\psi_N
=0 \, .
\eeq 
For $R > {\cal R}$, where $V({\cal R}) \ll E - N(N+1)\hbar^2/(2\mu{\cal 
R}^2)$, $\psi_N$ takes the approximate form (exact for $R \rightarrow 
\infty$ or $V \rightarrow 0$)
\beq
\label{eq:psil}
\psi_N \simeq {\cal S}_N(kR) \cos\delta_N + {\cal C}_N(kR) \sin\delta_N
\eeq
where $\delta_N$ is the phase shift of the $N^{\rm th}$ partial wave, and 
${\cal S}_N(kR)$ and ${\cal C}_N(kR)$ are Riccati-Bessel functions, which 
are
related to the spherical Bessel and spherical Neumann functions by 
${\cal S}_N(x)=x j_N(x)$ and ${\cal C}_N(x)=-x n_{N}(x)$.  Using 
Eq.~(\ref{eq:psil}) and the asymptotic forms
\beq
\lim_{R \rightarrow \infty}{\cal S}_N(kR) \rightarrow \sin
\left(kR - \frac{N\pi}2\right)
\, 
{\rm ~and~} \, 
\lim_{R \rightarrow \infty}{\cal C}_N(kR) \rightarrow \cos
\left(kR - \frac{N\pi}2\right),
\eeq
it is straightforward to prove the \emph{exact} integral identities
\beq
\sin\delta_N=-\frac{2\mu}{\hbar^2}\int_{0}^{\infty} V(R) 
\psi_N(R){\cal S}_{N}(kR) \,\rmd R ,
\eeq
and
\beq
\cos\delta_N=\frac{2\mu}{\hbar^2}\int_{0}^{\infty} \left(V(R) 
+ \frac{2N}{R^2}\right) \psi_N(R){\cal S}_{N-1}(kR) \, \rmd R.
\eeq
We use these integral forms to solve for $\delta_N=\delta_N(k)$ while 
simultaneously solving Eq.~(\ref{eq:rad}) for $\psi_N(R)$.  Given these 
phase shifts the scattering amplitude,
\beq
f(\hat\bOmega) = \frac{1}{k} \sum_{N=0}^{\infty} 
(2N+1)e^{\rmi\delta_N}\sin\delta_N\; P_N(\hat\bk\cdot\hat\bOmega) \, ,
\eeq
can immediately be found.

For this calculation we needed the scattering amplitude in the singlet 
$X^1\Sigma_g^+$ and triplet $b^3\Sigma_u^+$ potentials for H-H scattering, 
and in the He-H $X^2\Sigma^+$ scattering potential.  We used the same H-H 
potentials used in \citet{2005astro.ph..5173S}, which is a combination of 
several previously published H$_2$ surfaces, interpolations, and 
asymptotic formulas \citep{kolos1967, 1986JChPh..84.3278K, 
1989JChPh..91.2366F, 1992PhRvA..46.6956J}.  For the He-H potential, we 
used the fitting formula by \citet{1990PhRvA..42..311T}.

\end{document}